\documentclass[twocolumn]{aastex63}

\usepackage[utf8]{inputenc}
\DeclareUnicodeCharacter{2212}{-}% support older LaTeX versions
\usepackage{graphicx}
\graphicspath{{./figures/}{./}}
\usepackage{amsmath}
\usepackage{afterpage}
\usepackage{enumitem}
\setenumerate{itemsep=0mm}%noitemsep}
\hypersetup{    
  urlcolor        = {blue},
}
\usepackage{comment}
\usepackage{multirow}

\usepackage{lipsum, babel}

\usepackage{float}

\usepackage{placeins}

\usepackage{rotating}

\usepackage{lineno}
\usepackage{soul} 
\usepackage{amsmath}
\usepackage{amssymb}
\usepackage{xspace}
\usepackage{xifthen}
\usepackage{eso-pic}

\definecolor{forestgreen}{HTML}{228B22}
\definecolor{urlblue}{HTML}{000000}

\newcommand{\CHECK}[1]{{#1}}

\newcommand{\Gaia}{{\it Gaia}\xspace}
\newcommand{\SSSSS}{${S}^5$\xspace}

\mathchardef\mhyphen="2D

\newcommand{\roughly}{\ensuremath{ {\sim}\,} }

\newlength{\dhatheight}

\newcommand{\code}[1]{\texttt{#1}\xspace}

\newcommand{\unit}[1]{\ensuremath{\mathrm{\,#1}}\xspace}

\newcommand{\Myr}{\unit{Myr}}

\newcommand{\asec}{\unit{arcsec}}

\newcommand{\km}{\unit{km}}
\newcommand{\kms}{\km \second^{-1}}

\newcommand{\kpc}{\unit{kpc}}
\newcommand{\second}{\unit{s}}

\newcommand{\Msun}{\unit{M_\odot}}

\newcommand{\magn}{\unit{mag}}

\newcommand{\e}{\unit{e^{-}}}

\newcommand{\secref}[1]{Section~\ref{sec:#1}}
\newcommand{\appref}[1]{Appendix~\ref{app:#1}}
\newcommand{\tabref}[1]{Table~\ref{tab:#1}}

\newcommand{\figref}[1]{Figure~\ref{fig:#1}}

\newcommand{\bandvar}[2][]{%
  \ifthenelse{\isempty{#1}}{\var{#2}}{\var{#2\_#1}}%
}

\newcommand{\LCDM}{\ensuremath{\rm \Lambda CDM}\xspace}

\newcommand{\var}[1]{\ensuremath{\texttt{\MakeUppercase{#1}}}\xspace}

\providecommand\physrep{\ref@jnl{Phys.~Rep.}}%
\providecommand\apjs{\ref@jnl{ApJS}}%
\providecommand{\jcap}{\ref@jnl{JCAP}}%
\begin{document}

% A comparative study of the...
\title{Streams on FIRE: \\Populations of Detectable Stellar Streams in the Milky Way and FIRE}

\author[0000-0003-2497-091X]{Nora~Shipp}
\affiliation{MIT Kavli Institute for Astrophysics and Space Research, 77 Massachusetts Ave., Cambridge, MA 02139, USA}

\author[0000-0001-5214-8822]{Nondh~Panithanpaisal}
\affiliation{Department of Physics \& Astronomy, University of Pennsylvania, 209 S 33rd St., Philadelphia, PA 19104, USA}

\author[0000-0003-2806-1414]{Lina~Necib}
\affiliation{MIT Kavli Institute for Astrophysics and Space Research, 77 Massachusetts Ave., Cambridge, MA 02139, USA}
\affiliation{Department of Physics, Massachusetts Institute of Technology, Cambridge, MA 02139, USA}

\author[0000-0003-3939-3297]{Robyn~Sanderson}
\affiliation{Department of Physics \& Astronomy, University of Pennsylvania, 209 S 33rd St., Philadelphia, PA 19104, USA}
\affiliation{Center for Computational Astrophysics, Flatiron Institute, 162 5th Ave., New York, NY 10010, USA}

\author[0000-0002-8448-5505]{Denis~Erkal}
\affiliation{Department of Physics, University of Surrey, Guildford GU2 7XH, UK}

\author[0000-0002-9110-6163]{Ting~S.~Li}
\affiliation{Department of Astronomy and Astrophysics, University of Toronto, 50 St. George Street, Toronto ON, M5S 3H4, Canada}

\author{Isaiah~B.~Santistevan}
\affiliation{Department of Physics and Astronomy, University of California, Davis, CA 95616, USA}

\author[0000-0003-0603-8942]{Andrew~Wetzel}
\affiliation{Department of Physics and Astronomy, University of California, Davis, CA 95616, USA}

%%%%%%%%%%%%%%%%%%%%%%%%%%%%%%%%
% BUILDERS

\author[0000-0001-8536-0547]{Lara~R.~Cullinane}
\affiliation{Department of Physics and Astronomy, Johns Hopkins University, 3400 N. Charles St, Baltimore, MD 21218, USA}

\author[0000-0002-4863-8842]{Alexander~P.~Ji}
\affiliation{Department of Astronomy \& Astrophysics, University of Chicago, 5640 S Ellis Avenue, Chicago, IL 60637, USA}
\affiliation{Kavli Institute for Cosmological Physics, University of Chicago, Chicago, IL 60637, USA}

\author[0000-0003-2644-135X]{Sergey~E.~Koposov}
\affiliation{Institute for Astronomy, University of Edinburgh, Royal Observatory, Blackford Hill, Edinburgh EH9 3HJ, UK}
\affiliation{Institute of Astronomy, University of Cambridge, Madingley Road, Cambridge CB3 0HA, UK}
\affiliation{Kavli Institute for Cosmology, University of Cambridge, Madingley Road, Cambridge CB3 0HA, UK}

\author[0000-0003-0120-0808]{Kyler~Kuehn}
\affiliation{Lowell Observatory, 1400 W Mars Hill Rd, Flagstaff,  AZ 86001, USA}
\affiliation{Australian Astronomical Optics, Faculty of Science and Engineering, Macquarie University, Macquarie Park, NSW 2113, Australia}

\author[0000-0003-3081-9319]{Geraint~F.~Lewis}
\affiliation{Sydney Institute for Astronomy, School of Physics, A28, The University of Sydney, NSW 2006, Australia}

\author[0000-0002-6021-8760]{Andrew~B.~Pace}
\affiliation{McWilliams Center for Cosmology, Carnegie Mellon University, 5000 Forbes Ave, Pittsburgh, PA 15213, USA}

\author[0000-0003-1124-8477]{Daniel~B.~Zucker}
\affiliation{School of Mathematical and Physical Sciences, Macquarie University, Sydney, NSW 2109, Australia}
\affiliation{Macquarie University Research Centre for Astronomy, Astrophysics \& Astrophotonics, Sydney, NSW 2109, Australia}

%%%%%%%%%%%%%%%%%%%%%%%%%%%%%%%%
% COMMENTERS

\author[0000-0001-7516-4016]{Joss~Bland-Hawthorn}
\affiliation{Sydney Institute for Astronomy, School of Physics, A28, The University of Sydney, NSW 2006, Australia}
\affiliation{Centre of Excellence for All-Sky Astrophysics in Three Dimensions (ASTRO 3D), Australia}

\author[0000-0002-6993-0826]{Emily~C.~Cunningham}
\affiliation{Center for Computational Astrophysics, Flatiron Institute, 162 5th Ave., New York, NY 10010, USA}

\author[0000-0001-7052-6647]{Stacy~Y.~Kim}
\affiliation{Department of Physics, University of Surrey, Guildford GU2 7XH, UK}

\author[0000-0001-9046-691X]{Sophia~Lilleengen}
\affiliation{Department of Physics, University of Surrey, Guildford GU2 7XH, UK}

\author[0000-0002-3430-3232]{Jorge~Moreno}
\affiliation{Department of Physics and Astronomy, Pomona College, Claremont, CA 91711, USA}
\affiliation{Downing College, University of Cambridge, Cambridge CB3 OHA, UK}

\author[0000-0002-0920-809X]{Sanjib~Sharma}
\affiliation{Sydney Institute for Astronomy, School of Physics, A28, The University of Sydney, NSW 2006, Australia}
\affiliation{Centre of Excellence for All-Sky Astrophysics in Three Dimensions (ASTRO 3D), Australia}

\collaboration{21}{(\SSSSS Collaboration \& FIRE Collaboration)}
\email{nshipp@mit.edu}

\begin{abstract}
We present the first detailed study comparing the populations of stellar streams in cosmological simulations to observed Milky Way dwarf galaxy streams.
In particular, we compare streams identified around Milky Way analogs in the FIRE-2 simulations to stellar streams observed by the Southern Stellar Stream Spectroscopic Survey (\SSSSS). 
For an accurate comparison between the stream populations, we produce mock Dark Energy Survey (DES) observations of the FIRE streams and estimate the detectability of their tidal tails and progenitors. 
The number and stellar mass distributions of detectable stellar streams is consistent between observations and simulations. 
However, there are discrepancies in the distributions of pericenters and apocenters, with the detectable FIRE streams, on average, forming at larger pericenters (out to $> 110 \kpc$) and surviving only at larger apocenters ($ \gtrsim 40 \kpc$) than those observed in the Milky Way. 
We find that the population of high-stellar mass dwarf galaxy streams in the Milky Way is incomplete. Interestingly, a large fraction of the FIRE streams would only be detected as satellites in DES-like observations, since their tidal tails are too low-surface brightness to be detectable.
We thus predict a population of yet-undetected tidal tails around Milky Way satellites, as well as a population of fully undetected low-surface brightness stellar streams, and estimate their detectability with the Rubin Observatory.
Finally, we discuss the causes and implications of the discrepancies between the stream populations in FIRE and the Milky Way, and explore future avenues for tests of satellite disruption in cosmological simulations.
\end{abstract}


\section{Introduction}
\label{sec:intro}

The $\Lambda$ Cold Dark Matter (\LCDM) cosmological model predicts that galaxies form via hierarchical merger events and accretion. Within this framework, galaxies like the Milky Way reside within dark matter halos, which are built up via the accretion and disruption of lower-mass subhalos. Surviving subhalos may be traced by observations of the luminous satellite galaxies that they host; and subhalos undergoing tidal disruption can be traced by the tidal remnants of these systems -- stellar streams. 

Observations of surviving satellite galaxies have enabled strong constraints on near-field cosmology, including unprecedented insight into the intricacies of galaxy formation at small scales and the properties of dark matter, raising important challenges to the \LCDM cosmological model \citep{Bullock:2017}, such as the missing satellites \citep{Moore:1999, Klypin:1999}, core-cusp \citep{Navarro:1996}, and too-big-to-fail \citep{Boylan-Kolchin:2011, Boylan-Kolchin:2012} problems. These controversies have in large part been resolved by incorporating the effect of baryonic physics into cosmological simulations \citep[e.g.][]{Wetzel:2016, Garrison-Kimmel:2017, Fitts:2017, Simpson:2018, Garrison-Kimmel:2019, Buck:2019, Kim:2019, Sales:2022}, and by the discovery of a large number of Milky Way satellites in wide-area optical imaging surveys \citep[see][and references therein]{Drlica-Wagner:2020}. Despite this progress, however, many open questions remain. The nature of dark matter is one of the largest outstanding questions in modern physics, and we have yet to build a comprehensive understanding of galaxy formation at the smallest scales.

Stellar streams are strong complementary probes of near-field cosmology and have the power to provide additional tests of our understanding of galaxy formation and dark matter in the local universe. Satellite tidal disruption remains a large source of uncertainty in studies of near-field cosmology \citep[e.g.][]{Carlsten:2020}. Incorporating stellar streams into studies of Milky Way and simulated satellite populations enables comparisons of not only the surviving population of satellites, but of their rate of disruption. Since tidal disruption is sensitive to the density profiles of satellites, which in turn are highly dependent on properties of the dark matter particle \citep[e.g.][]{Tulin:2018, Du:2018} as well as baryonic physics \citep[e.g.][]{Garrison-Kimmel:2019}, constraints on disruption rates in the Milky Way may be be used to test theories of both dark matter and galaxy evolution \citep[e.g.][]{Penarrubia:2012, Errani:2015}.

Stellar streams also enable precise measurements of the local gravitational potential \citep[e.g.][]{Johnston:2005, Law:2010, Bovy:2014, Gibbons:2014}, including the overall mass and shape of the Milky Way halo and its massive satellites \citep{Erkal:2019, Vasiliev:2021, Shipp:2021}. These measurements place the Milky Way in a cosmological context, enabling more precise tests of galaxy formation and dark matter physics with cosmological simulations. In addition, streams are one of a small number of methods predicted to be able to detect the presence of low-mass subhalos that host no luminous baryonic component \citep[][]{Johnston:2002, Ibata:2002, Carlberg:2009, Koposov:2010, Yoon:2011, Carlberg:2012, Erkal:2015b, Sanders:2016, Bovy:2017, Price-Whelan:2018, Bonaca:2018b}.

Streams have traditionally been difficult to incorporate into population-level comparisons of satellites in simulations and observations. However, with wide-area surveys, the number and quality of observations of stellar streams have increased dramatically in recent years. To date, nearly 100 tidal remnants of dwarf galaxies and globular clusters have been discovered around the Milky Way \citep{Mateu:2022}. In addition, for the first time, kinematic measurements are available for a large population of stellar streams, thanks to proper motions measured by \Gaia \citep{Gaia:2016, Gaia:2018, Gaia:2021}, as well as radial velocities obtained by large-scale spectroscopic surveys of stellar streams \citep[e.g.][]{Li:2019, Zhao:2012, Majewski:2017}. At the same time, zoom-in cosmological simulations \citep[e.g.][]{Wetzel:2016, Hopkins:2018} are now able to resolve dwarf galaxy streams (down to $M_* \gtrsim 5 \times 10^5 \Msun$) around Milky Way-mass hosts \citep{Panithanpaisal:2021}. Comparing such populations of Milky Way and simulated stellar streams will enable a broad range of tests of hierarchical structure formation, tidal disruption, and the nature of dark matter. 

\citet{Li:2022} presented an overview of a population of one dozen stellar streams with complete 6D phase space measurements, observed by the Southern Stellar Stream Spectroscopic Survey \citep[\SSSSS;][]{Li:2019}. In comparing these Milky Way streams to those found in the Latte suite of simulations \citep{Wetzel:2016}, based on FIRE-2 physics \citep{Hopkins:2018}, they raised the possibility of a ``too-big-to-fail'' problem in stellar streams. In particular, they found an  excess of high-stellar mass streams ($M_* \gtrsim 5 \times 10^5 \Msun$) in FIRE relative to the population observed in the Milky Way. While the FIRE Milky Way analogs have a median of 8 stellar streams in this mass range, only $\roughly 2$ have been observed around the Milky Way, perhaps suggesting that a population of high-stellar mass, low-surface brightness streams remain undetected in the Milky Way. Alternatively, this discrepancy between simulations and observations may imply that FIRE is over-disrupting or otherwise over-producing massive streams.%which will have implications for tests of \LCDM with both intact and disrupting satellite galaxies.

The FIRE simulations have been found to reasonably reproduce the population of surviving satellites observed around the Milky Way, including the distributions of stellar masses and velocity dispersions \citep{Wetzel:2016, Garrison-Kimmel:2019}, as well as the radial distance distribution \citep{Samuel:2020}. Similar comparisons of populations of tidally disrupting satellites further test the agreement between simulations and observations, enabling stronger constraints on and new insight into dark matter and galaxy formation physics in the local universe. %the implementation of dark matter and galaxy formation physics in cosmological simulations.

In this paper, we compare detectable stellar streams in the FIRE simulations to dwarf galaxy streams observed in the Milky Way. We produce mock Dark Energy Survey \citep[DES;][]{DES:2005, DES:2018, DES:2021} observations of the FIRE streams to estimate their detectability. We then compare the number, stellar mass, and orbital distributions of these populations. We find that when taking detectability into account, the number and stellar mass distribution of these streams are consistent between simulations and observations. However, the orbital distributions of these populations differ significantly, with FIRE streams existing at large pericenters and apocenters relative to those in the Milky Way. We present our observed Milky Way and simulated FIRE stream datasets in \secref{data}; in \secref{mocks} we discuss our mock observations and estimate the detectability of stellar streams and satellites in FIRE. In \secref{results} we compare the populations of stellar streams in the Milky Way and in FIRE; and in \secref{disc} we discuss the discrepancies between these two populations, the implications for satellite galaxy disruption in the Milky Way, and predictions for future observations.

\section{Simulations \& Data}
\label{sec:data}
\subsection{Stellar Streams in the Milky Way}
\label{sec:obs}

In this paper, we focus on the dwarf galaxy streams that have been detected around the Milky Way.\footnote{The Latte and ELVIS on FIRE simulations used in this work do not include globular clusters at $z =0$, thus we only consider dwarf galaxy streams in this analysis.}
The population of known streams in the Milky Way has increased dramatically in recent years, thanks to the advent of large photometric and astrometric surveys \citep{Grillmair:2016, Shipp:2018, Malhan:2018b, Mateu:2018}. The total number of known streams is approaching 100 \citep{Mateu:2022}, and will continue to grow with ongoing and upcoming surveys \citep[e.g.][]{Drlica-Wagner:2021, LSST:2009}. Recent \Gaia data releases \citep{Gaia:2021, Gaia:2018} have enabled measurements of the proper motions of stellar streams out to $\roughly 50 \kpc$ \citep[e.g.][]{Price-Whelan:2018, Malhan:2018b, Shipp:2019, Koposov:2019, Ramos:2020, Ibata:2020}, providing the first kinematic measurements of a large population of stellar streams.

In addition, \SSSSS has obtained spectroscopic measurements of $>20$ stellar streams, using the Two-degree Field fiber positioner \citep{Lewis:2002} and the AAOmega spetrograph \citep{Sharp:2006} on the 3.9m Anglo-Australian Telescope. The metallicity and radial velocity measurements obtained by \SSSSS enable classification of stream progenitors, metallicity-based estimates of total stellar mass, and precise orbit modeling. 

\citet{Li:2022} summarized the orbital and chemical properties of the first dozen \SSSSS stellar streams. Of the 12 streams, six are determined to be disrupted dwarf galaxies (OC, Elqui, Indus, Palca, Turranbura, and Jhelum). These six streams are classified as dwarf galaxy streams due to their resolved calcium triplet (CaT) metallicity dispersions and large radial velocity dispersions ($\sigma_{\rm [Fe/H]} > 0.2\ \mathrm{and/or}\ \sigma_{\rm v} > 10 \kms$), relative to the globular cluster streams. In addition to these six streams, we include in our analysis the Sagittarius stream \citep[Sgr;][]{Ibata:1994, Majewski:2003}, which has been observed by \SSSSS as well as by several others \citep{Hasselquist:2019, Yang:2019, Johnson:2020}. These seven streams make up the population of confirmed dwarf galaxy streams around the Milky Way.\footnote{Although these streams all lie within the DES footprint, they make up the full current sample of known Milky Way dwarf galaxy streams. Cetus and LMS-1 are also likely dwarf galaxy streams; however, the former is likely associated with the Palca stream \citep{Li:2022, Yuan:2022}, and the latter is connected to the Indus stream \citep{Malhan:2021}.}

In order to compare these stellar streams to those identified in FIRE, we determine the stellar masses and orbits of each system. As described in \citet{Li:2022}, the total luminosities of these streams are calculated from the measured metallicities, using the empirical relation from \citet{Simon:2019}. The luminosities are then converted to stellar mass, assuming $M_*/L_{\rm V} = 1.6$ \citep{Kirby:2013}. The scatter in this relation is 0.16 dex in [Fe/H], which corresponds to a factor of 3.5 uncertainty in stellar mass. This method provides measurements of the total stellar mass, which do not rely on the detection of the full extent of each stream.

Incorporating the redshift dependence of the mass-metallicity relation, as discussed, for example, in \citet{Naidu:2022} (N22), would increase the resulting stellar masses of the Milky Way streams, reducing the scale of the discrepancy between the Milky Way and FIRE. However, we choose not to use the mass-metallicity relation for disrupting systems derived in N22 for several reasons. First, the relation is fit primarily to phase-mixed systems, and the two streams that are considered (Sgr and OC) are outliers from the resulting fit. Second, the systems considered in N22 have large stellar masses ($M_* > 10^{6.1} \Msun$), and extrapolating the relation to smaller masses would overestimate the masses of faint satellites that are likely to be quenched before infall. In addition, in FIRE, we see that surviving satellites and streams follow the same mass-metallicity relation \citep{Panithanpaisal:2021}. Using the N22 relation effectively increases the stellar masses of the Milky Way streams by a factor of ten. This leads to an excess of low mass streams in the Milky Way relative to those seen in FIRE. At the higher mass end, the stellar mass distributions of detectable streams remain consistent within $1 \sigma$. We therefore use the $z=0$ relation throughout this work, and leave a determination of the mass-metallicity relation for Milky Way dwarf galaxy streams to future work.

The orbit of each \SSSSS stream is determined by fitting the radial velocities from \SSSSS, proper motions from \Gaia EDR3, and distances of blue horizontal branch and RR Lyrae member stars in the best-fit static Milky Way potential from \citet{McMillan:2017}, as described in \citet{Li:2022}. The pericenters and apocenters reported in \citet{Li:2022} are compared to the models of \citet{Shipp:2021}, which use a modified Lagrange Cloud Stripping technique \citep{Gibbons:2014} to model stream disruption in the presence of the Milky Way and Large Magellanic Cloud (LMC). The resulting orbits are found to be consistent, and therefore the values presented in Table 2 of \citet{Li:2022} are used throughout this paper.

\begin{figure*}[tph]
    \centering
    \includegraphics[width=0.8\textwidth]{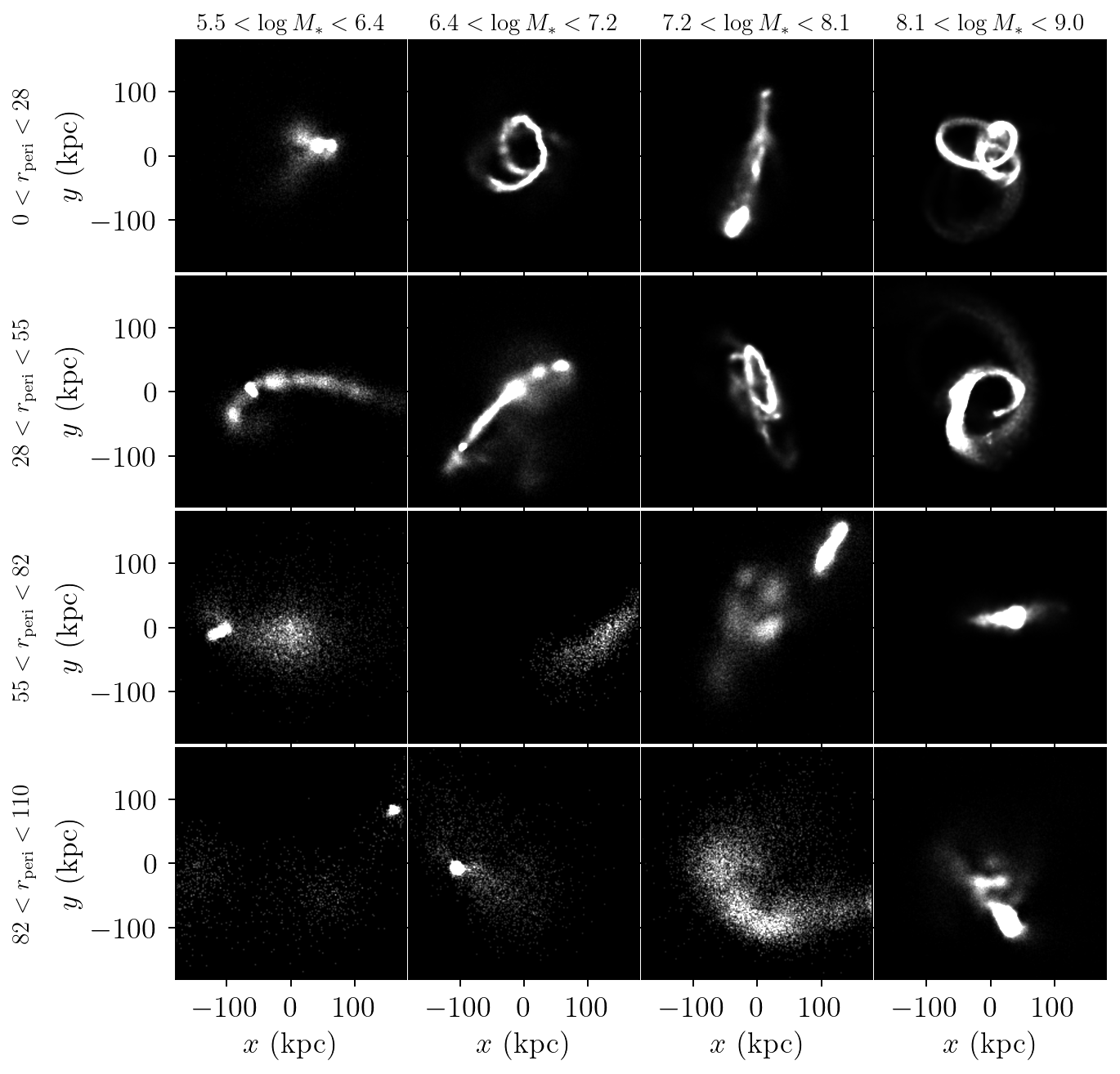}
    \caption{A selection of simulated stellar streams in FIRE. Synthetic stars are sampled from the simulation star particles, as described in \secref{mocks}. The figure columns are grouped by stellar mass (in units of $\Msun$) and the rows are grouped by pericenter, to illustrate a subset of the simulated stream population with a range of parameters and morphologies. Each panel shows the 2D density of synthetic stars in coordinates centered on the host galaxy. While all streams do have coherent, extended structures, some, particularly those with large pericenters, have very low-surface brightness tidal tails.}
    \label{fig:stream_grid}
\end{figure*}

\subsection{Stellar Streams in FIRE}
\label{sec:sims}

Zoom-in cosmological-baryonic simulations of galaxy formation can now achieve resolutions that allow for the study of tidal remnants of dwarf galaxies around Milky Way-like hosts, produced self-consistently via accretion from the cosmic web. In this work, we focus on one such set of simulated Milky Way analogs: the FIRE-2 ``Latte'' and ``ELVIS on FIRE'' suites \citep{Wetzel:2016,Hopkins:2018,Garrison-Kimmel:2019,Samuel:2020}. 
The FIRE-2 simulations are run with the \texttt{Gizmo}\footnote{\url{https://bitbucket.org/phopkins/gizmo-public}} code \citep{Hopkins:2013,Hopkins:2015}, which uses a mesh-free finite-mass Lagrangian Godunov method for the hydrodynamic solver, and a version of the Tree-PM solver based on \texttt{GADGET-3} \citep{Springel:2005}. Star formation and feedback are implemented using the FIRE-2 prescriptions described in \citet{Hopkins:2018}.

These simulations produce galaxies with properties that are similar, but not identical, to those of the Milky Way in many aspects, including the stellar mass and structure of the central disk \citep{Sanderson:2020} and the structure of the stellar halo \citep{Sanderson:2018}. The population of satellite galaxies around each host also resembles the Milky Way's in key aspects, including their mass function, size-mass relation, and internal velocity dispersion \citep{Wetzel:2016}, radial distribution \citep{Samuel:2020}, star formation histories \citep{Garrison-Kimmel:2019b}, and mass-metallicity trends \citep{Panithanpaisal:2021}. The fidelity of the dwarf satellite population supports our use of these simulations to represent the population of tidally disrupted dwarfs, which form a different subset of the same accreted population \citep{Panithanpaisal:2021, Cunningham:2021}. 

In this work, we focus on the stellar streams first identified by \cite{Panithanpaisal:2021} in zoom-in simulations of the seven isolated\footnote{No equally massive halo within 10 Mpc.} Milky Way-like galaxies from the Latte suite \citep{Wetzel:2016}, and three Milky Way+M31-like pairs from the ELVIS on FIRE suite \citep{Garrison-Kimmel:2019}, for a total of 13 host galaxies. These galaxies have halo masses at $z=0$ of $M_{\rm 200} = 1 \mhyphen 2.1 \times 10^{12} \Msun$ \citep{Sanderson:2018}. The Latte suite has an initial stellar mass resolution of $7070 \Msun$ per star/gas particle, and includes the simulations m12i, m12f, m12m, m12b, m12c, m12w, and m12r. %m12i, first introduced in \citep{Wetzel:2016}, has a relatively quiet merger history, while m12f \citep{Garrison-Kimmel:2017} has a late active merger, as analyzed in \citep{Necib19}. 
The ELVIS on FIRE suite includes the pairs Thelma and Louise, Romeo and Juliet, and Remus and Romulus. The initial stellar mass resolution of these galaxies is $3500 \mhyphen 4000 \Msun$ per particle.  \cite{Sanderson:2018} discusses the detailed differences in the stellar halos of each of these galaxies. A table of the main properties of the central galaxies is available in \citet{Santistevan:2020, Bellardini:2021}, and the satellite population is further discussed in \citet{Garrison-Kimmel:2019} and \citet{Samuel:2020}.

\citet{Panithanpaisal:2021} identified streams around these halos by tracking luminous substructures that were bound and within the $z=0$ virial radius of the host ($\roughly 350 \kpc$) at any time between 2.7--6.5 Gyr ago\footnote{Systems that become unbound earlier than 6.5 Gyr ago have been classified as phase-mixed at $z=0$ and are discussed in greater detail in Horta et al. (in prep.). Systems accreted more recently are predominately on first infall and have not had sufficient time to disrupt into coherent streams.}. The stream candidates are then selected using the following criteria: First, they must have between 120 and $10^5$ star particles. This translates to stellar masses between $\roughly 5 \times 10^{5} \Msun$ and $\roughly 10^9 \Msun$. Second, the maximum pairwise distance between member star particles must be greater than $120 \kpc$, indicating that the stream has stretched at least partway across the galaxy.\footnote{As discussed in \citet{Panithanpaisal:2021}, this constraint effectively eliminates dwarf galaxies that are somewhat tidally deformed but do not have coherent tidal tails. Removing this constraint increases the sample of coherent streams, as identified by eye, by one across the 13 simulations.} Third, the stream candidates must have a local velocity dispersion below a stellar mass-dependent threshold (see their Equation 2), requiring that they remain coherent in phase space. We note that the accreted debris in FIRE can have complex morphologies (particularly at the massive end; e.g., Cunningham et al, in prep.), with some portions of the debris remaining kinematically coherent and other portions fully phase-mixing. However, for the purposes of this study, and for consistency of the selection, we limit ourselves to systems that are classified as streams by the selection criteria of \citet{Panithanpaisal:2021}.

In order to ensure accurate measurements of the stellar masses and orbits of the FIRE streams, we update the selection of member stars beyond what has been done in \citet{Panithanpaisal:2021}. We require that each stream member star is associated with the progenitor for at least ten snapshots; this eliminates contamination from unassociated satellites and from host disk stars that may be erroneously picked up by the halo finder due to coincidental proximity in phase space. This updated selection changes the number and the stellar masses of the streams presented in \citet{Panithanpaisal:2021}. 

We then compute the orbits of the FIRE streams by individually tracing the orbits of each member star through the saved snapshots, and then integrating the orbit around each pericenter using the code \code{AGAMA} \citep{Vasiliev:2019}. We integrate each star particle in the corresponding host galaxy potential at the snapshot closest to the most recent pericentric passage\footnote{The time frequency of the output snapshots (600 snapshots, spaced by $\roughly 10 \Myr$) is high enough that the integration between snapshots generally has very little effect on the resulting pericentric distance.}, using the models from \citet{Arora:2022}\footnote{\href{https://web.sas.upenn.edu/dynamics/data/pot_models/}{https://web.sas.upenn.edu/dynamics/data/pot\_models/}}. These models are fit to each snapshot, and consist of multipole expansions (up to $l=4$), computed using a combination of spherical harmonics and azimuthal harmonics for the halo and disk respectively. We then fit a Gaussian kernel density estimate (KDE) to the spatial distribution of stars at $z=0$ to select the highest density region, corresponding to the portion of the stream around the surviving or dissolved progenitor. We take the median pericenter and apocenter of stars in this high-density region as the overall pericenter and apocenter of the stellar stream. This procedure excludes very diffuse stream components and produces measurements of the orbital parameters more consistent with how we measure stream orbits in the Milky Way.

The properties of the updated stream population are summarized in \tabref{streams} in \appref{sys_props}.

\section{Mock Observations}
\label{sec:mocks}

\begin{figure*}[htp]
    \centering
    \includegraphics[width=0.99\textwidth]{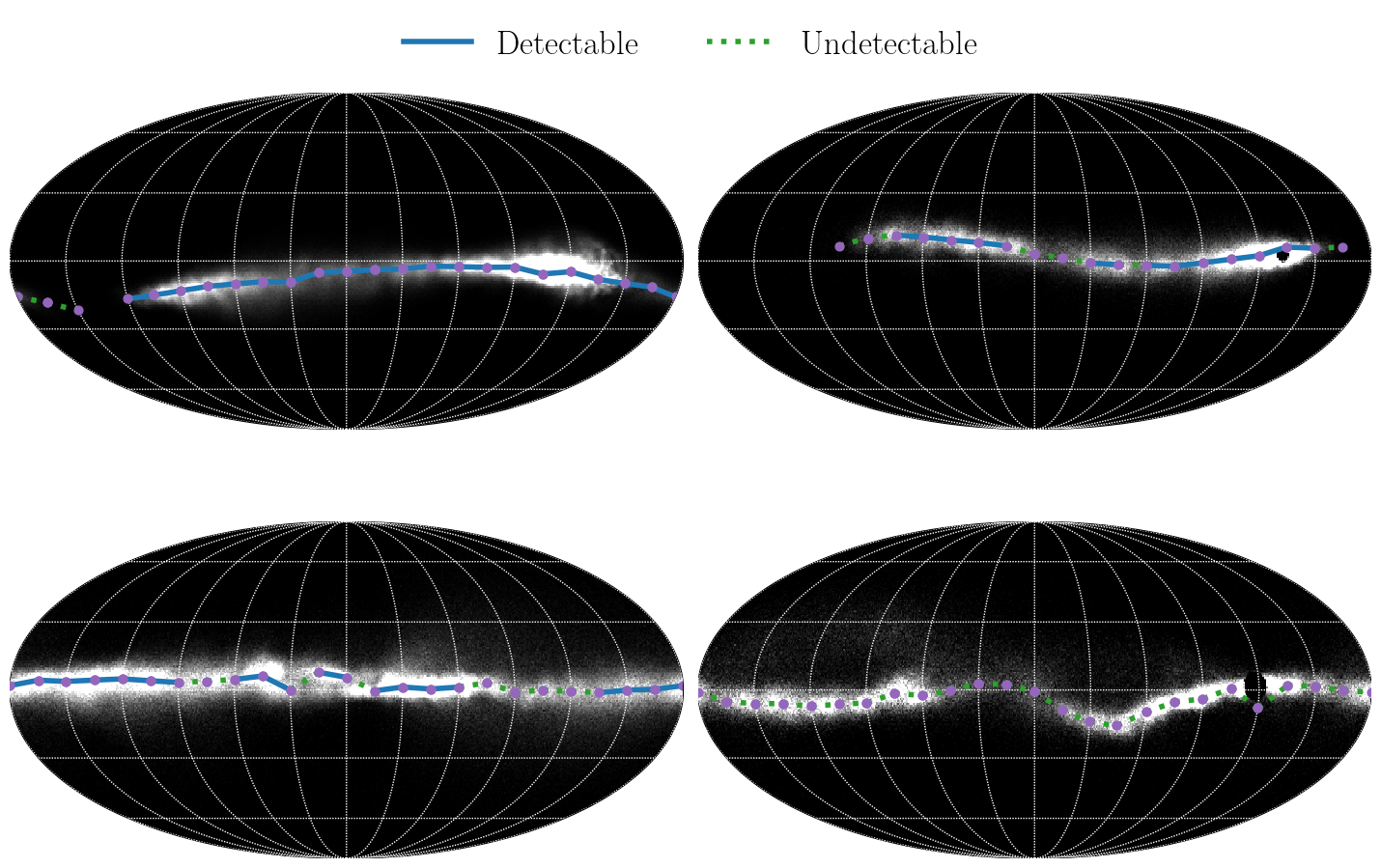}
    \caption{Detectability of a sample of simulated stellar streams. Each panel shows the mock observation of one stream, transformed into stream coordinates (so that the stream lies along $\phi_{\rm 2} \roughly 0$). In order to estimate detectability, we divide each stream into $15 \deg$ segments, and calculate the average surface brightness within the $1 \sigma$ Gaussian width. Blue segments are detectable ($\mu_{\rm V} < 34\ \mathrm{mag/arcsec^2}$), and green dashed segments are undetectable. Surviving progenitors are masked within $4 r_{1/2}$. The underlying image shows the number density of stars along the stream.}
    \label{fig:stream_det}
\end{figure*}

In order to compare more directly to observed Milky Way stellar streams, we produce mock DES observations of the FIRE streams and estimate the detectability of these systems. The DES data is well-suited to the study of stellar streams; its precise photometry and depth of observations have enabled the discovery and detailed study of a large population of stellar streams \citep{Shipp:2018}. The DES streams are generally lower surface brightness and more distant than other known streams. In addition, the majority of streams that have been followed up by \SSSSS, which make up the population of Milky Way stellar streams considered in this work, have been observed with DES. The DES footprint covers $\roughly 5000 \deg^2$, one eighth of the sky. Similar photometric surveys, such as the Sloan Digital Sky Survey \citep[SDSS; ][]{York:2000}, Pan-STARRS \citep{Chambers:2016}, the DESI Legacy Imaging Surveys \citep{Dey:2018}, and the DECam Local Volume Exploration Survey \citep[DELVE; ][]{Drlica-Wagner:2021} span much of the remaining area, albeit to a shallower magnitude depth. Estimating the detectability of the FIRE streams in all-sky DES photometry will therefore only overestimate the total detectable number, and thereby provide a conservative test of whether considering detectability is sufficient to resolve the too-big-to-fail problem in stellar streams.

We use the \code{Ananke} code \citep{Sanderson:2020} to simulate the mock observations. We generate a population of synthetic stars from each simulated star particle. The simulation particles represent the average position, velocity, age, and metallicity of an ensemble of stars. For the Latte simulations, the initial mass of each gas particle is $7070 \Msun$; at $z=0$ the average mass of the resulting star particles is $\roughly 5000 \Msun$. The ELVIS simulations have star particles masses of $\roughly 3.5 \mhyphen 4 \times 10^3 \Msun$. As in \citet{Sanderson:2020, Sharma:2011}, we consider each star particle to represent a population of stars with a single age and metallicity. We sample individual stellar masses from a Kroupa initial mass function (IMF) \citep{Kroupa:2001}. We then sample absolute magnitudes in the DES $grizY$ bands from an isochrone \citep{Bressan:2012} with the age and metallicity of the parent star particle.

For each host galaxy, we establish a coordinate system centered on a solar viewpoint and local standard of rest (LSR), with a solar position within the host disk plane at a Galactocentric distance of $R_{\odot} = 8.2 \kpc$ \citep{Bland-Hawthorn:2016}. We perform a rotation around the z-axis such that solar position lies in the -$x$ direction. We then transform into the LSR frame using the solar position and solar velocity from \code{Astropy}\citep{astropy:2013,astropy:2018, astropy:2022}.

We place each synthetic star in position and velocity space, by sampling an Epanechikov kernel \citep{Epan:1969} centered on the parent particle. We consider position and velocity space independently, and the size of each kernel is inversely proportional to the cube-root of the local density around each parent particle. The local density is calculated from the 16 nearest neighbors, using the density estimator \code{EnBid} \citep{SharmaSteinmetz:2011}. The kernel shape was selected to be computationally efficient for massive streams, and the number of nearest neighbors was selected to ensure an accurate local density estimation for low-mass streams.

We then compute the apparent magnitude of each star, based on the assigned heliocentric distance, and convolve the apparent magnitude with the DES photometric uncertainties. We parameterize the $g$-band magnitude error as 

\begin{align}\begin{split}
err(g) =  0.0006 + e^{(g - 26.0) / 0.88},
\end{split}\end{align}

\noindent where the coefficients are fit to the DES DR2 \citep{DES:2021} weighted-average magnitudes and magnitude errors (\var{WAVG\_MAG\_PSF\_G, WAVG\_MAGERR\_PSF\_G}). We similarly parameterize the $r$ and $i$ band errors with equation 1, with coefficients (0.0002, 25.7, 0.87) and (0.0020, 25.1, 0.81), respectively. The $z$ and $Y$ band magnitudes are not used in this analysis.  When calculating the magnitude uncertainty for each synthetic star, we assume an interstellar reddening, $E(B - V)$, of 0.04, roughly the average across the DES footprint.

The resulting dataset for each mock-observed stellar stream is available for download at \href{https://flathub.flatironinstitute.org/sapfire}{https://flathub.flatironinstitute.org/sapfire}. \figref{stream_grid} displays a selection of these mock-observed streams, binned by stellar mass (columns) and pericenter (rows) to show a broad sample of FIRE stream morphologies.

\subsection{Stream Detectability}
\label{sec:stream_det}

We determine the detectability of each FIRE stream following a procedure motivated by matched-filter stream searches in the DES data \citep{Shipp:2018}. The detectability of a stellar stream can be approximated based on the average surface brightness of observed member stars (i.e. stars falling within the DES magnitude limits of $16 < g < 24.7$). Lower luminosity and more diffuse stellar streams (i.e. fainter surface brightness) are more difficult to disentangle from contaminating foreground and background stellar populations. This method also accounts for the fact that more distant streams will have fewer detectable stars, and will therefore have a fainter observed surface brightness. \citet{Shipp:2018} published the average surface brightness of the observed portion of each of the DES stellar streams. We correct these values to include only stars within the DES magnitude limits, and find that the surface brightness limit for stellar streams detectable within DES is $\mu_{\rm V} \leq 34 \magn/\asec^2$.

In order to calculate the average surface brightness of the tidal tails of each FIRE stream, we first mask all stars within $4 r_{1/2}$ of the progenitor, using the progenitor parameters derived below (\secref{sat_det}). We then convert to a coordinate system where the stream lies approximately along the equatorial plane, where $\phi_1$ is the coordinate along the stream and $\phi_2$ is the coordinate perpendicular to the stream. 
In practice, stars belonging to a long, convoluted stream do not share the exact same orbital plane, so we estimate the best-fit RA and Dec of the orbital pole $(\alpha, \delta)$ for each stream
by minimizing the quantity $\mathcal{D}(\alpha, \delta) = \sum_i |\frac{\pi}{2} - \mathcal{D}_i(\alpha, \delta)|$, where $\mathcal{D}_i(\alpha, \delta)$ is the angular distance from the pole to each stream member star.

We then divide the stream into equal-length ($15 \deg$) segments and individually calculate the detectability of each segment to account for variation in surface brightness along the stream. Then, to determine the $\phi_2$ position of each endpoint, we take stars that are within 3 degrees of $\phi_1^{\rm end}$. If there are fewer than 50 stars in this region, we disregard this endpoint as it is likely an empty space, and automatically mark its neighboring segments as undetectable. Otherwise, we determine the corresponding $\phi_2^{\rm end}$ to be the peak in the $\phi_2$ distribution of the selected stars. We determine the detectability of each of these segments individually, and consider streams to be detectable if they have at least one detectable segment. We count all systems with $>1$ detectable segment as one detectable stream (just as we consider associated, but disconnected, segments of known Milky Way dwarf galaxy streams to be one single stream).

Within each segment, we perform another coordinate transformation so that the great circle connecting the two end points lies along the equatorial plane. We then determine the width in $\phi_{\rm 2}$ of the segment by fitting the peak interval (i.e. the smallest $\pm 1 \sigma$ interval containing the peak of the distribution). We consider only the inner $\pm 1 \sigma$ range in order to exclude diffuse components at large $\phi_2$ that would bias the detectability estimate. We then select all stars within this range, and convert their DES $g$ and $r$ band magnitudes to visual magnitudes using the relation from \citet{Bechtol:2015},

\begin{align}\begin{split}
V = g_{\rm DES} - 0.487(g_{\rm DES} - r_{\rm DES}) - 0.025.
\end{split}\end{align}

We then compute the total luminosity ($M_{\rm V}$) and area of each segment, and calculate the average surface brightness  ($\mu_{\rm V}$). \figref{stream_det} illustrates the detectability of segments along 4 of the FIRE streams. The detectable (blue) and undetectable (green) segments are overplotted on the number density of stars along the stream.

\begin{figure*}[htp]
    \centering
    \includegraphics[width=0.9\textwidth]{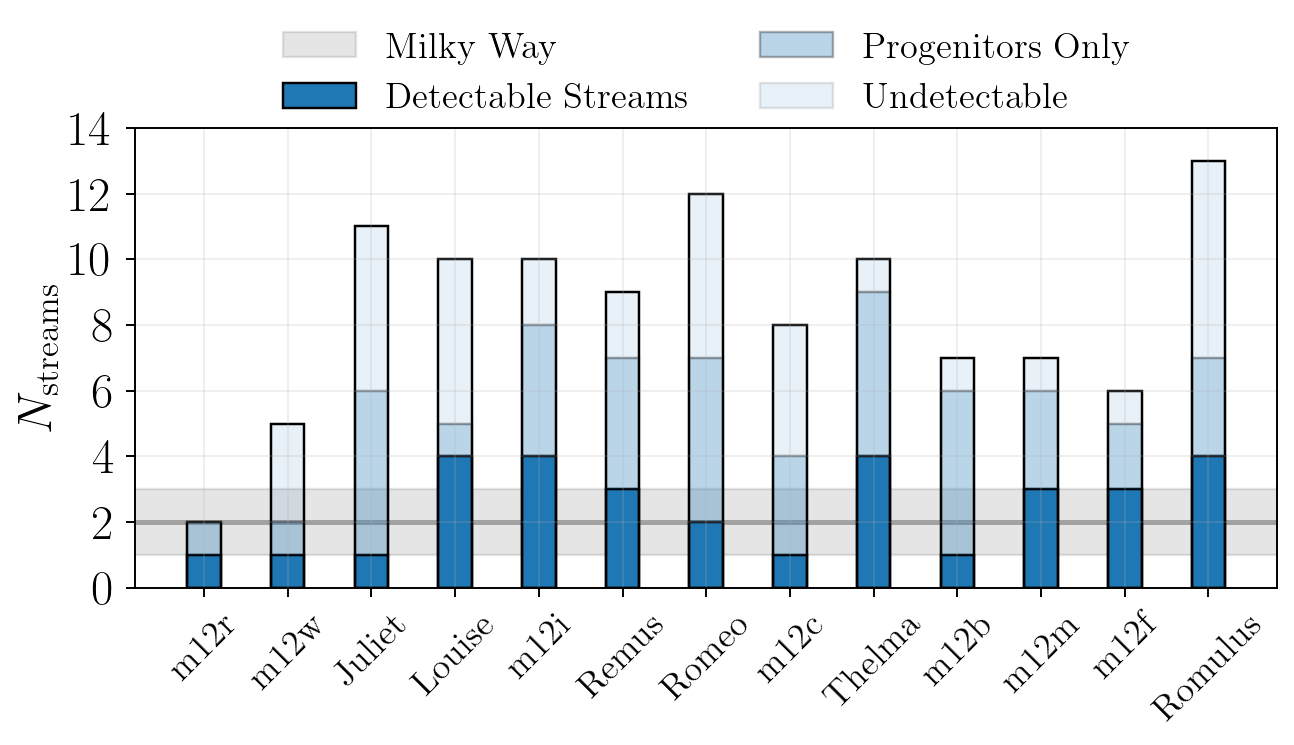}
    \caption{Total and detectable stream populations around 13 FIRE hosts. The bottom segment of each bar represents the number of streams with detectable tidal tails, the middle segment represents the number of streams with only detectable progenitors (i.e. would be mistaken as intact satellites), and the top segment indicates the remaining number of streams. The total height of each bar is the total stream population in each galaxy. The host halos are sorted in order of host halo mass (low to high). The gray band represents the $1 \sigma$ scatter in the number of streams, within the stellar mass range considered, that have been detected within the Milky Way (the scatter is due to uncertainty in stellar mass measurements). The number of detectable streams in FIRE is consistent with the number detected in the Milky Way.}
    \label{fig:det}
\end{figure*}

We find that \CHECK{32 of the 96} streams with $M_* \gtrsim 5 \times 10^5 \Msun$ are detectable across the 13 FIRE halos. This corresponds to a median of \CHECK{3} streams per host galaxy. The number of detectable streams per host is shown in \figref{det}, and these results are discussed in greater detail in \secref{num}.

\begin{figure}[htp]
    \centering
    \includegraphics[width=0.95\columnwidth]{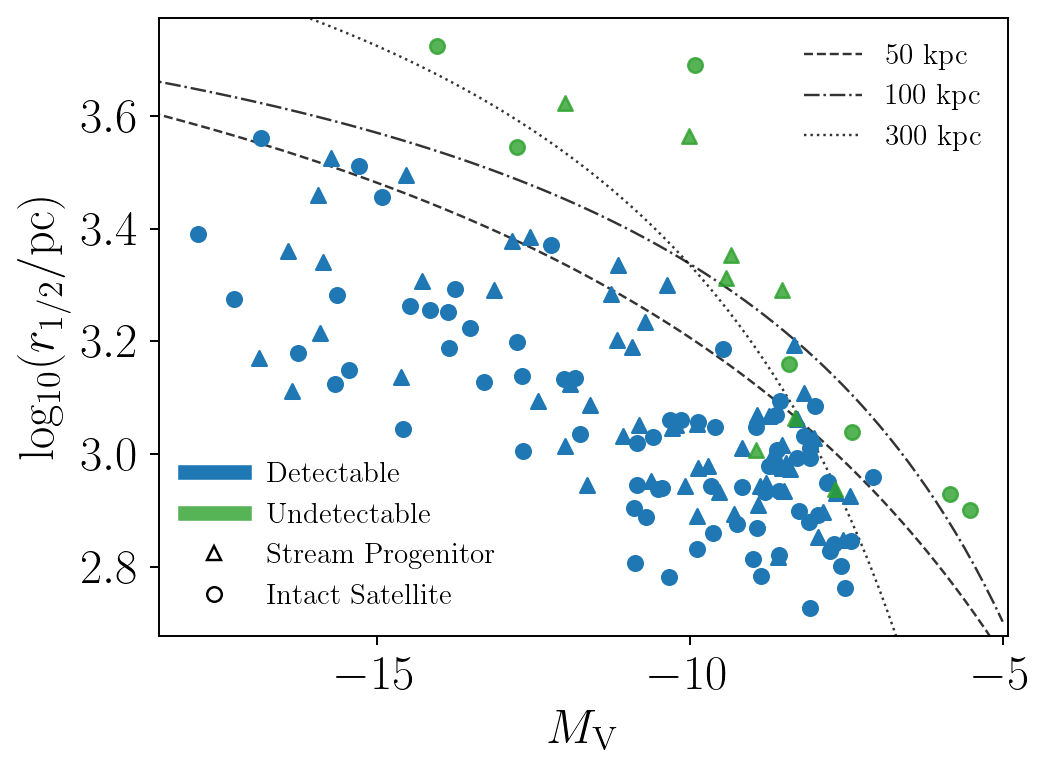}
    \caption{Estimated satellite detectability, using the analytic approximation from \citet{Drlica-Wagner:2020}. Luminosity, half-light radius, and distance  at $z=0$ are derived from the mock observations described in \secref{sat_det}. Triangles are surviving stream progenitors, and circles are intact satellites. The majority of satellites in this mass and distance range are detectable (blue), whereas only a small number of surviving satellites and stream progenitors are undetectable (green). The curves represent $50\%$ detectability at different distances.}
    \label{fig:sat_det}
\end{figure}

\subsection{Satellite Detectability}
\label{sec:sat_det}

We also estimate the detectability of intact satellites and stream progenitors in FIRE. Gravitationally-bound dark matter halos and subhalos are identified in the simulations and assigned dark matter, gas, and star particles using the \code{ROCKSTAR} halo finder \citep{Behroozi:2013c}, as described in \citet{Wetzel:2016}. We select satellites from the halo catalog at $z=0$ with more than $ 100$ star particles ($M_{*} \gtrsim 5 \times 10^5 \Msun$), $d < 300 \kpc$, and bound mass fraction $> 0.4$. Across the 13 host halos, this is a total of \CHECK{140} satellites, \CHECK{61} of which are stream progenitors (a median of $55\%$ and minimum of $40\%$ of streams in this mass range have surviving progenitors, i.e. have progenitors that are recovered by the halo finder at $z=0$). Throughout this paper, we refer to satellites included in this sample as ``surviving satellites.'' 
Among the surviving satellites, we refer to satellites with no evidence of tidal disruption as ``intact satellites'' and those with tidal tails as ``stream progenitors.''

We simulate the mock DES observations of these satellites, similarly to the streams, as described in \secref{mocks}. We then use a procedure analogous to the studies of observed satellite galaxies \citep[e.g.][]{Bechtol:2015, Drlica-Wagner:2015, Drlica-Wagner:2020} to fit an elliptical \citet{Plummer:1911} profile to each mock-observed satellite. We use the code \code{ugali}\footnote{\href{https://github.com/DarkEnergySurvey/ugali}{https://github.com/DarkEnergySurvey/ugali}} \citep{Bechtol:2015, Drlica-Wagner:2015} to measure the half-light radius ($r_{1/2}$), ellipticity ($\epsilon$), position angle ($\theta$), and centroid ($\alpha_0, \delta_0$) of each system. We calculate the total luminosity ($M_{\rm V}$) by converting the mock DES $g$-band magnitudes to visual magnitudes as in \citet{Bechtol:2015}, and correcting for the unobserved luminosity (due to DES magnitude limits) using the synthetic ischrones from \citet{Bressan:2012} implemented in \code{ugali}. We take the distance to each satellite to be the median distance of mock-observed member stars. The derived properties of each satellite are listed in \tabref{sats} in \appref{sys_props}.

\citet{Drlica-Wagner:2020} presented an analytic approximation of the detectability of satellite galaxies as a function of absolute magnitude, half-light radius, and distance. We use the derived parameters described above to estimate satellite detectability (\figref{sat_det}). We find that the majority of satellites in this distance and stellar mass range, including the majority of surviving stream progenitors, would be detectable with DES, with \CHECK{72 out of 79} intact satellites and \CHECK{53 out of 61} stream progenitors detectable. One important caveat is that this approximation is based on dwarf galaxy search techniques that are optimized for ultra-faint dwarf galaxies ($M_{\rm V} \gtrsim -7.7$), and may therefore underestimate the detectability of the brightest satellites. However, we find on average $< 1$ satellite per host within this luminosity range is undetectable. This assumption therefore has a negligible effect on our overall conclusions. We combine these results with the stream detectability estimates in \secref{stream_det} in order to identify the number of streams with detectable tidal tails, detectable progenitors only (i.e. would be mistaken as intact satellites), or no detectable component (\figref{det}). The population of streams mistaken as intact satellites is discussed in greater detail in \secref{disc}. %The satellite properties are summarized in \tabref{sats}.  
For clarification of the terms used throughout this paper, \tabref{class} lists the resulting classification of stream and satellite systems, depending on the detectability of their progenitors and tidal tails.

\section{Comparison between Milky Way and FIRE Stellar Streams}
\label{sec:results}

\subsection{Number of Streams}
\label{sec:num}

This work was motivated in part by the question of whether there is an excess of high-stellar mass streams in the FIRE simulations relative to what we observe in the Milky Way: i.e. is there a too-big-to-fail problem in stellar streams? In order to address this question, we produced mock observations of the FIRE streams and estimated their detectability. We can now make a consistent comparison between the population of Milky Way streams and those in FIRE.

In \figref{det}, we show the number of detectable streams around each FIRE halo in comparison to the number of streams observed in this stellar mass range in the Milky Way. The gray line represents the number of Milky Way dwarf galaxy streams. We find that the Milky Way has $2 \pm 1$ of these luminous systems (Sgr, OC, and Jhelum), where the spread is due to the uncertainties in the stellar mass measurements (i.e. within uncertainties, both OC and Jhelum could be below the stellar mass limit considered here). Each bar represents the number of streams in one of the simulated Milky Way analogs in FIRE. The bottom segment is the number of streams with detectable tidal tails, the middle segment is the number with only a detectable progenitor (i.e. they would be mistaken as an intact satellite galaxy), and the top segment represents the number of undetectable streams. The full height of the bar is the total number of streams in this stellar mass range in each halo. Here we include only streams with $M_* \gtrsim 5 \times 10^5 \Msun$ for consistency, although the ELVIS simulations are slightly higher resolution and have streams down to $M_* \gtrsim 3 \times 10^5 \Msun$. The bars are sorted in order of increasing host halo mass. Within this mass range, we see no meaningful correlation between host halo mass and number of streams. Without taking into account detectability, the number of streams in FIRE would be inconsistent with what we have observed in the Milky Way. However, we find that many streams, even at this luminosity, remain undetectable due to their low surface brightness. We also find that many detectable satellite galaxies have undetectable tidal tails. Therefore, the observed Milky Way streams are consistent with the detectable FIRE streams and there is no ``too-big-to-fail'' problem in stellar streams. \textit{In addition, the FIRE simulations predict a population of yet-undetectable massive streams and a population of satellites with significant, but still undetected, tidal tails.} We discuss the implications of this unobserved population of streams, and make predictions for the future detectability of these systems in \secref{disc}.

\begin{figure*}[htp]
    \centering
    \includegraphics[width=0.99\textwidth]{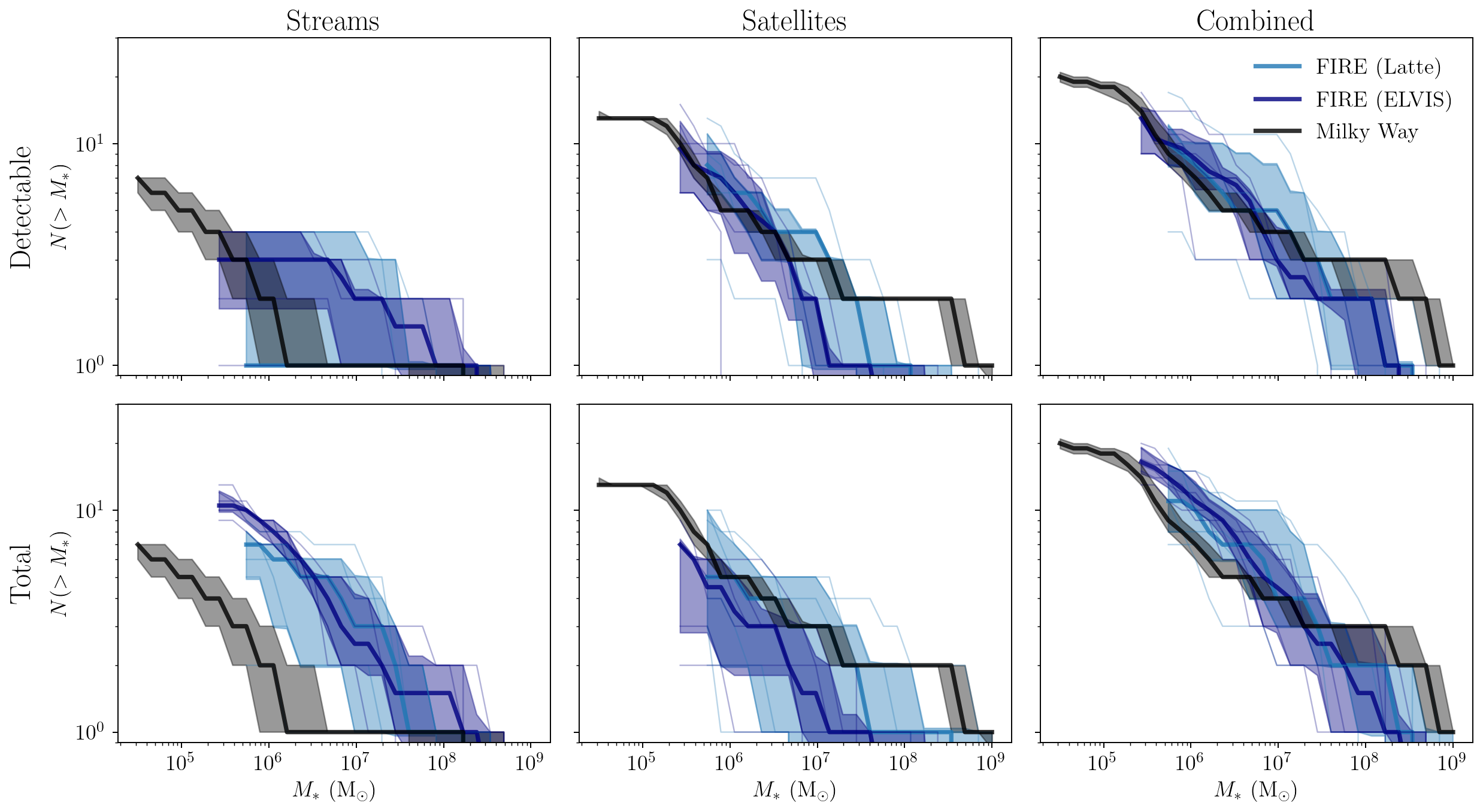}
    \caption{Cumulative mass distribution of streams and satellites in FIRE and the Milky Way. The upper row includes only detectable systems, as determined in \secref{mocks}. The lower row includes all systems in FIRE. The left column includes only streams, the central only intact satellites (no tidal tails/no detectable tidal tails), and the right includes the combination of these two populations. The uncertainty on the Milky Way curves (black) comes from scatter in the stellar mass-metallicity relations used to estimate the total stellar mass of each system. The shaded bands on the FIRE curves represents the scatter between systems. When taking into account detectability, the FIRE stream and satellite mass distributions are largely consistent with those observed in the Milky Way. The ELVIS halos have a slight excess of streams with $M_* > 10^6 \Msun$ relative to the Milky Way, however the discrepancy is only $\sim 1 \sigma$.}
    \label{fig:stellar_mass}
\end{figure*}

\subsection{Stellar Mass}
\label{sec:mass}

In addition to the total number, we consider the stellar mass distribution of streams and satellites in FIRE and in the Milky Way. In \figref{stellar_mass} we present the cumulative stellar mass distribution. The top row includes only detectable systems, and the bottom row includes all systems in FIRE. The Milky Way systems are the same between the two rows. In all panels, the black curve represents the Milky Way systems and the blue curves represent FIRE. We plot the Latte (light blue) and ELVIS (dark blue) streams independently, because the simulations have slightly different resolution limits, and to explore possible differences between stream populations in paired and isolated Milky Way-like systems. The shaded bands around the blue curves represents the scatter between systems. The uncertainty on the Milky Way stream stellar masses corresponds to the scatter in the stellar mass-metallicity relation, as discussed in \secref{data}. The Milky Way satellite stellar masses are calculated from luminosities compiled by \citet{Putman:2021, McConnachie:2012}, and stellar mass-to-light ratios from \citet{Woo:2008}. As uncertainties, we take the typical 0.17 dex uncertainty on stellar mass reported by \citet{Woo:2008}.

In each row, the left panel includes only stellar streams, the middle panel includes only intact satellites (no tidal tails), and the right panel includes the combined population of streams and intact satellites. The Milky Way is largely consistent with the detectable systems in FIRE, with the exception of the LMC at the high-mass end.\footnote{None of the Latte or ELVIS on FIRE galaxies have LMC analogs surviving to $z=0$, though some have analogous accretion events in the past \citep{Samuel:2021}. We leave a detailed study of the effect of the accretion histories of simulated halos on their stream populations to future work.}
In addition, the ELVIS pairs have a larger number of streams with $M_* > 10^6 \Msun$, leading to a $\roughly 1 \sigma$ discrepancy in the stellar mass distribution in this range. This is consistent with the host halos in the Local Group-like pairs having formed earlier, as discussed in \citet{Garrison-Kimmel:2019, Santistevan:2020}. 

The lower-middle panel includes only intact satellites, with no tidal tails. The upper-middle panel includes observable satellites with no detectable tails, including streams that would be mistaken as intact satellites. We find that many surviving satellites have tidal tails. This suggests that for full consistency, comparisons between surviving satellites in FIRE and in the Milky Way should include dwarf galaxy stellar streams with surviving progenitors (e.g. Sgr).

\begin{figure*}[thp]
    \centering
    \includegraphics[width=0.95\textwidth]{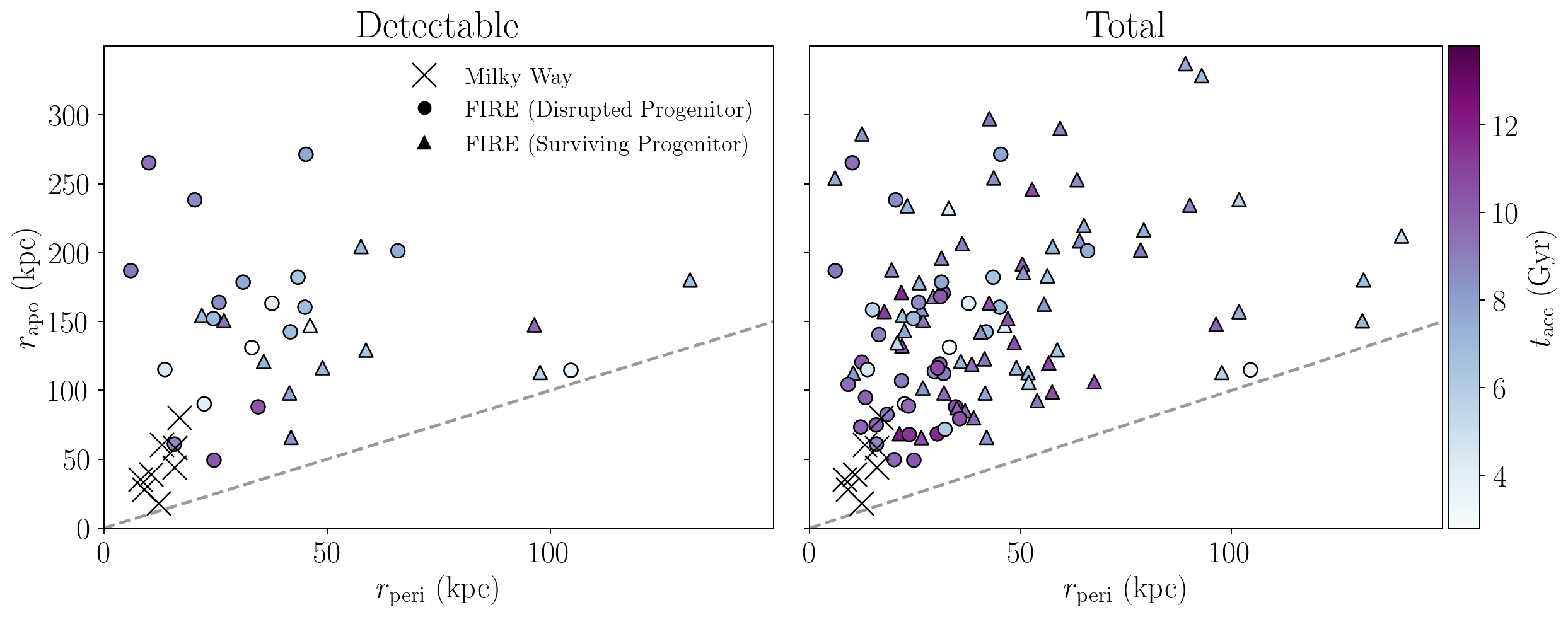}
    \caption{Most recent pericenters and apocenters of the FIRE and Milky Way stellar streams. The black crosses indicate the Milky Way dwarf galaxy streams, where $r_{\rm peri}$ and $r_{\rm apo}$ are taken from \citet{Li:2022}. The circles represent the FIRE streams with fully disrupted progenitors, and the triangles indicate FIRE streams with surviving progenitors. Among the Milky Way streams, Sgr (largest pericenter and apocenter) is the only one with a known surviving progenitor. The color indicates lookback time since accretion. The left panel includes only detectable streams, and the right panel includes the full population of streams in FIRE. The dashed line indicates a circular orbit (equal pericenter and apocenter).}
    \label{fig:peri_apo}
\end{figure*}

\begin{figure}[htp]
    \centering
    \includegraphics[width=0.9\columnwidth]{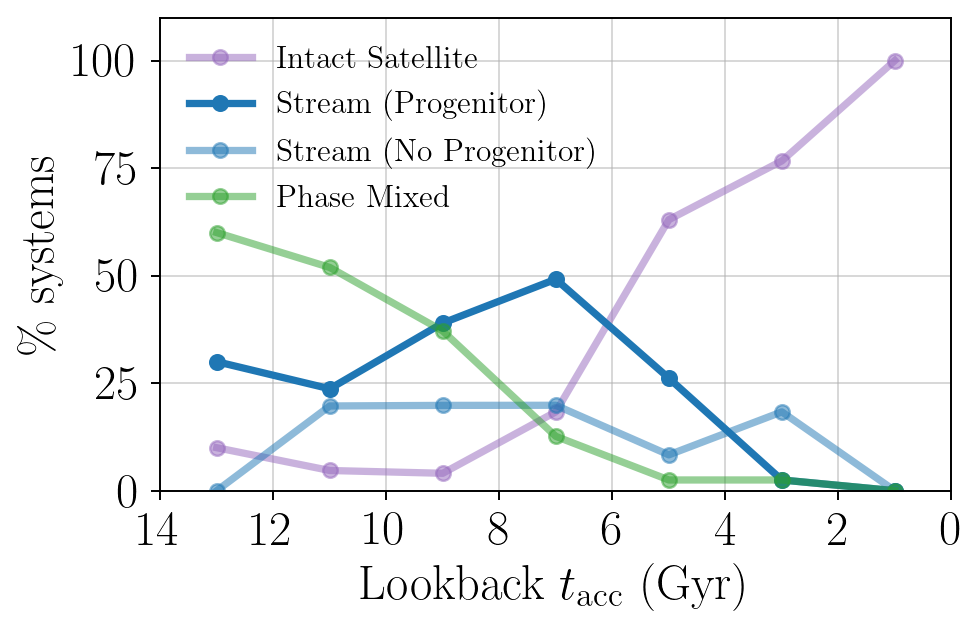}
    \caption{Frequency of systems as a function of lookback time since accretion. Each line indicates the percentage of systems in each time bin classified as an intact satellite (purple), stream with a surviving progenitor (dark blue), stream without a surviving progenitor (light blue), or phase-mixed (green). The plotted values are the means of the percentages across the 13 halos.}
    \label{fig:accretion_time}
\end{figure}

\begin{figure}[htp]
    \centering
    \includegraphics[width=0.9\columnwidth]{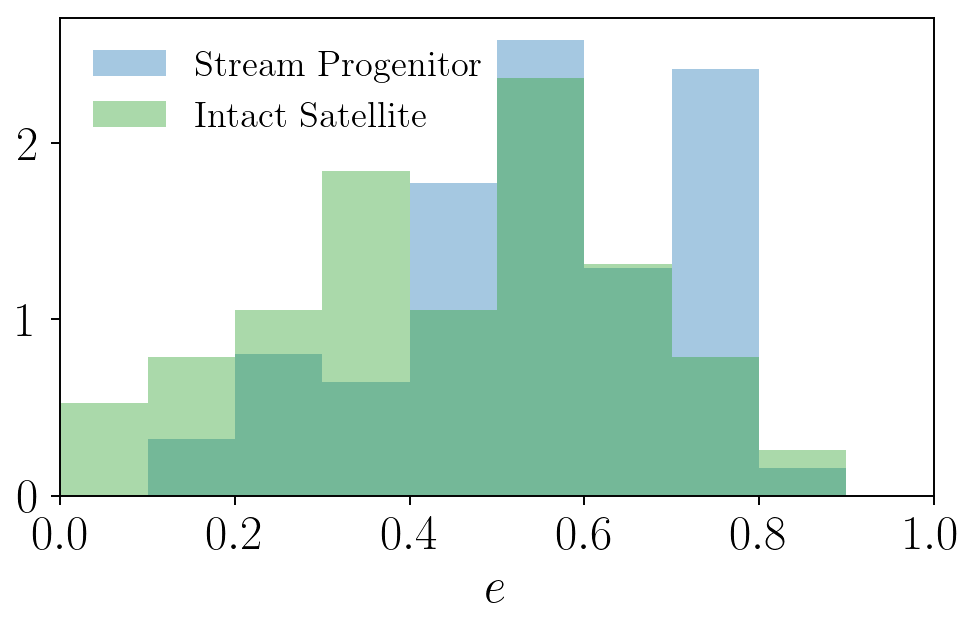}
    \caption{Orbital eccentricity of satellites in FIRE. The blue histogram includes satellites with tidal tails, classified as stellar streams, and the green histogram includes only satellites with no evidence of tidal disruption. As expected, tidally disrupting satellites are on slightly more radial orbits than intact satellites.}
    \label{fig:eccentricity}
\end{figure}

\begin{figure}[htp]
    \centering
    \includegraphics[width=0.9\columnwidth]{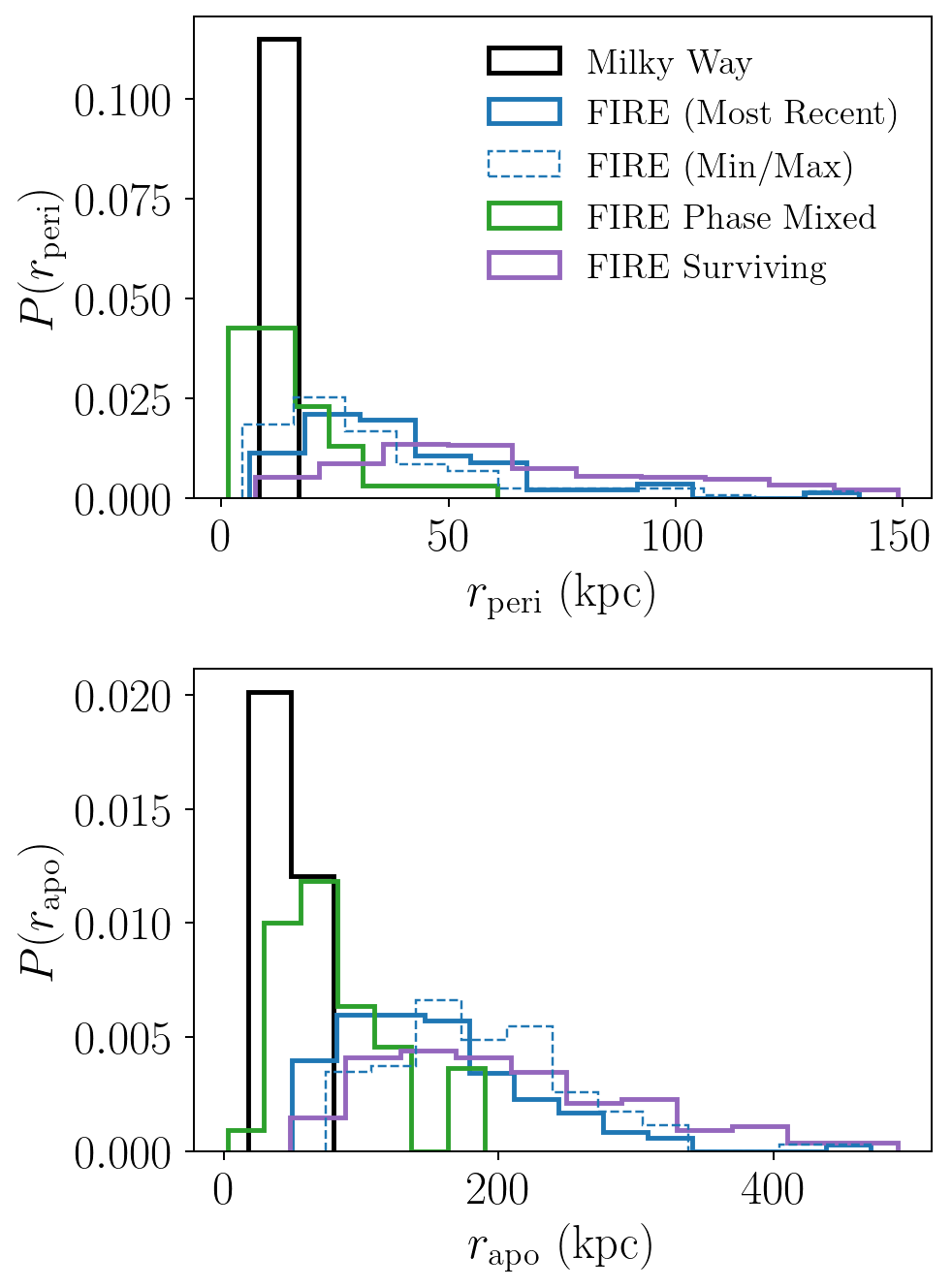}
    \caption{Distributions of pericenters (upper) and apocenters (lower) of systems in FIRE and the Milky Way. The black histograms indicate the Milky Way dwarf galaxy streams. The solid blue line represents the most recent pericenter and apocenter of the FIRE streams, which is most analogous to the values measured for the Milky Way. For comparison, we plot the distributions of the overall minimum (maximum) pericenter (apocenter) across all passages with the dashed blue lines. The purple histograms represent surviving satellites in FIRE (with or without tidal tails), and green represents phase-mixed systems.}
    \label{fig:peri_apo_hists}
\end{figure}

\subsection{Orbits}
\label{sec:orbits}

Finally, we consider the orbits of stellar streams in FIRE and the Milky Way. \figref{peri_apo} shows the pericenters and apocenters of the FIRE streams (with and without surviving progenitors), in comparison to the Milky Way dwarf galaxy streams. The left panel shows only the detectable FIRE streams and the right panel shows the full population of streams. Here, we plot the most recent pericenter and apocenter, for consistency with the values obtained by fitting orbits of Milky Way stellar streams. The FIRE streams are colored by lookback time at accretion. The triangles indicate streams with surviving progenitors, and circles are streams with fully disrupted progenitors.

We find that the distributions of pericenters and apocenters of streams in the Milky Way and FIRE are largely inconsistent, even once taking detectability into account. Many FIRE satellites are disrupting to form streams at large pericenters (out to $> 100 \kpc$), in contrast to the Milky Way streams, which have pericenters within $30 \kpc$. In addition, we find that the FIRE streams have relatively large apocenters, all $> 40 \kpc$. The FIRE streams with fully disrupted progenitors (circles) tend to have smaller pericenters ($r_{\rm peri} \lesssim 60 \kpc$) than those with surviving progenitors (triangles). This is more consistent with the Milky Way streams, of which only Sgr has a surviving progenitor. However, we note that streams with surviving progenitors tend to have detectable tidal tails only at smaller distances (and correspondingly, smaller apocenters) due to the relatively small fraction of stellar mass in their tidal tails.

The method of fitting orbits to the Milky Way streams is discussed in \secref{obs} and to the FIRE streams in \secref{sims}. The primary difference between the two methods is that we calculate the orbits of the FIRE streams by tracing the positions of stars back through the saved simulation snapshots, thereby accounting for the full time-dependence of the host galaxy potential. However, we integrate the orbits of the Milky Way streams in a time-independent potential. We examine the implications of this assumption in greater detail in \appref{orbit_comp}. We find that the resulting uncertainties are not sufficient to account for the qualitative difference observed between the pericenters and apocenters of streams in FIRE and the Milky Way.

A small number of FIRE streams do overlap with the Milky Way population. The four streams on the most similar orbits to the Milky Way streams are not tightly clustered in stellar mass or accretion time, but are all from the m12c and m12i simulations. These two galaxies have a relatively large number of streams and, on average, streams with smaller pericenters and apocenters, than the other simulated systems. %They also have relatively quiet merger histories (no accretion events with $M_* > 10^9 \Msun$ at $z < 1$), perhaps suggesting a relationship between stream populations and accretion history. 
However, even these two simulations have a majority of streams with larger pericenters and apocenters than those seen in the Milky Way. 

In addition, there is some difference between the orbits of streams in the Latte (isolated Milky Ways) and ELVIS (Local Group pairs) simulations. In particular, the streams with the largest pericenters are from the ELVIS simulations. Interestingly, \citet{Samuel:2020} found that the population of intact satellites in the Milky Way is, on average, more concentrated in radial distance than those in FIRE. Perhaps some feature of the Milky Way, such as its unusual assembly history \citep{Evans:2020} or its disk size, has led to a more radially concentrated population of satellites at all stages of disruption. Additional simulations, as well as observations of stream populations around external galaxies \citep[e.g.][]{Carlin:2016, Pearson:2019}, will enable studies of the effect of host halo properties and environment on populations of stellar streams.

We note that the detectable streams, on average, were accreted more recently than the undetectable streams. Many of the early-accreted, undetectable streams have relatively small pericenters, but are undetectable due to their large widths and low surface brightnesses. On the other hand, many of the undetectable streams with later accretion times have large apocenters, and are undetectable primarily due to their large distances.

\citet{Li:2022} found that more streams are on prograde orbits than retrograde orbits, relative to the Galactic disk in the Milky Way (only two out of the 12 studied are on retrograde orbits). Among the FIRE halos, we find a range of stream orbital orientations. Some have an excess of prograde or retrograde streams, while some do not. The effect of host properties, particularly accretion history, on the orbital distributions of stream populations in cosmological simulations will be explored in greater detail in future work.

\figref{accretion_time} shows the accretion times of systems (detectable and undetectable) at different stages of disruption in FIRE. The purple line is intact satellites with no tidal tails, dark blue is streams with a surviving progenitor, light blue is streams without a surviving progenitor, and green is phase-mixed systems. As expected, intact satellites dominate the population at recent accretion times and phase-mixed systems dominate at early accretion times, and stellar streams lie in between. There is no clear difference between the accretion times of streams with and without surviving progenitors, suggesting that orbit, rather than time since accretion, primarily determines the survival of the stream progenitor. Similarly, we find differences in orbits between satellites with and without tidal tails. \figref{eccentricity} shows the distributions of orbital eccentricities of intact and disrupting satellites. Those that are classified as stellar streams have, on average, a larger orbital eccentricity, suggesting that systems on more radial orbits are more likely to form tidal tails.

\figref{peri_apo_hists} shows a comparison of the distributions of pericenters and apocenters of these different systems. In black are the Milky Way dwarf galaxy streams. In solid blue are the FIRE streams, as in \figref{peri_apo}. The dashed blue lines are the largest (smallest) among all pericenters (apocenters) of each of the FIRE streams. Even the overall minimum pericenters are largely inconsistent with those measured in the Milky Way. In purple are the surviving satellites in FIRE (with or without tidal tails). The satellite orbits are presented in Santistevan et al. (in prep.). The pericenters are calculated by searching for local minima in satellite distance from their Milky Way-mass host in an adaptive time window, during the time after the satellite has first crossed the virial radius of the host. The authors then fit a cubic spline within a large enough time window, and save the minima of the spline-interpolated distances to ensure that these pericenters are real, and not artificial due to noise.
In green are the orbits of the phase-mixed systems in FIRE. Phase-mixed systems are classified as systems failing the median velocity dispersion criterion for stream classification, as described in \secref{sims}. These pericenters and apocenters are calculated in the same way as for the FIRE streams, described above. However, for these systems, we take the median of all stars, without first selecting the highest density regions. We find that the phase-mixed systems in FIRE have pericenters and apocenters that are most consistent with those measured for the Milky Way dwarf galaxy streams. Therefore, although the number and mass distribution of streams in the Milky Way and FIRE are consistent, when taking into consideration detectability, the orbits are still inconsistent, and may be suggesting that FIRE is disrupting and phase-mixing satellites at a higher rate than our own Galaxy. We discuss the implications of this discrepancy and plans for future studies of tidal disruption in FIRE in \secref{disc}.

\section{Discussion}
\label{sec:disc}

We find that, when taking into account detectability, the number and stellar mass distributions of streams around Milky Way analogs in the FIRE simulations are consistent with observations. However, we find a discrepancy in the distributions of orbital parameters. In particular, FIRE streams disrupt at a larger range of pericenters than observed in the Milky Way, and survive only at larger apocenters. Despite this disagreement, it is valuable to discuss the predicted population of undetectable streams. In the future, these predictions will either highlight further conflict between simulations and observations, or facilitate the continued discovery of stellar streams. Here, we discuss the implications of the predicted population of undetectable high-stellar mass streams, predictions for future stream discovery in the Milky Way, and possible explanations for the discrepancies between the stream populations in the Milky Way and FIRE.

\subsection{Undetected Streams}

We find that \CHECK{64 out of 96} stellar streams, across the 13 halos, would be undetected. The median number of undetectable streams per halo, in this stellar mass range, is \CHECK{5 out of 8}. In addition, we find that \CHECK{42 out of 64} stream progenitors (median \CHECK{3 out of 5}) would be mistaken for intact satellites given current observations with deep photometric surveys. %However, we note that complementary observations, such as proper motions from \Gaia and spectroscopy from \SSSSS, have enabled the detection of low surface brightness tidal features around satellites \citep{Ji:2021}.

\begin{figure}[tp]
    \centering
    \includegraphics[width=0.99\columnwidth]{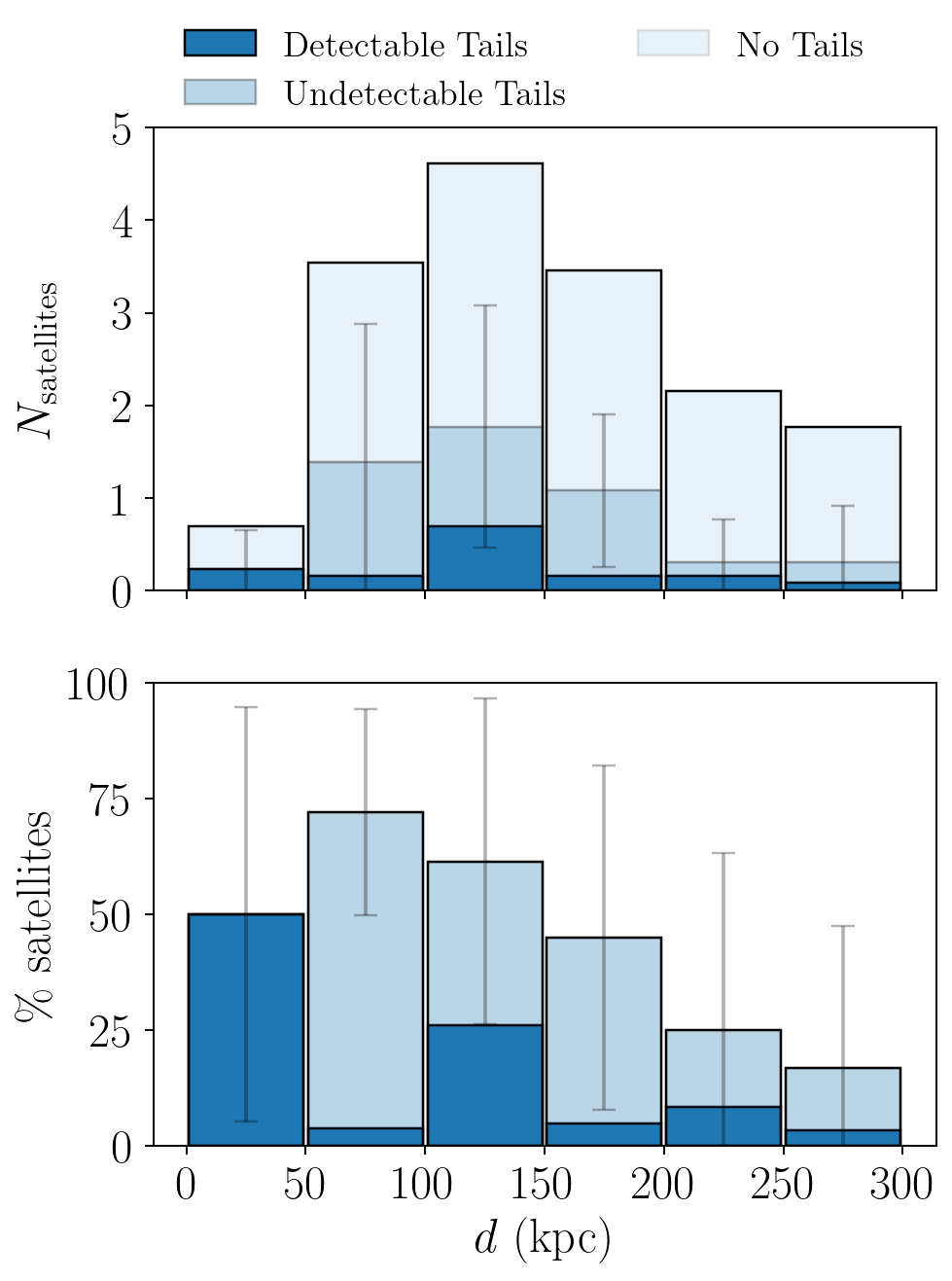}
    \caption{\textit{Upper:} Mean number of satellites with detectable tidal tails (dark blue), undetectable tidal tails (medium blue), and no tidal tails (light blue), binned by distance from the host galaxy. \textit{Lower:} Mean percentage of satellites with detectable (dark blue) and undetectable (medium blue) tidal tails. The remaining percentage are the satellites that are fully intact and have no tidal tails. These values are the mean across the 13 FIRE halos of each ratio, rather than the ratio of the means plotted above. The error bars represent the $1 \sigma$ scatter between halos in the total number or percentage of satellites with tidal tails (detectable or undetectable) in each distance bin.}
    \label{fig:distance_fraction}
\end{figure}

The FIRE simulations, if accurately reproducing tidal disruption in Milky Way-like galaxies, therefore suggest that the population of high-stellar mass Milky Way streams is incomplete, and that an undetected population of high-stellar mass, yet low-surface brightness, streams may exist in the Milky Way halo. In addition, they suggest that many systems we have classified as intact satellites may in fact be undergoing tidal disruption and have yet-undetected tidal tails.

\figref{distance_fraction} shows the distribution of satellites that have detectable and undetectable tidal tails as a function of distance at $z=0$ from the host galaxy. The upper panel shows the number of satellites with detectable, undetectable, or no tidal tails, and the bottom panel shows the percentage of satellites in each distance bin with detectable or undetectable tails. We find that at distances of $50 \mhyphen 200 \kpc$, more than $50 \%$ of satellites have tidal tails, many of which remain undetectable.

Several Milky Way satellites do show signs of tidal disruption. Among the stream sample ($M_* \gtrsim 5 \times 10^5 \Msun$) considered here, Sgr is the only one with a known surviving progenitor. Below this stellar mass range, the Tucana III stellar stream \citep{Drlica-Wagner:2015, Shipp:2018} also has a surviving progenitor with two extending tidal tails. More ambiguous are the Milky Way satellites without clear tidal tails, but that otherwise show evidence of tidal disruption. For example, Antlia II has a large size, low density, and a velocity gradient consistent with predictions of tidal disruption \citep{Torrealba:2019, Ji:2021, Vivas:2022}. With a stellar mass of $\roughly 10^6 \Msun$ and a distance of $\sim 130 \kpc$ \citep{Ji:2021}, Antlia II would fall within the $50 \mhyphen 100 \kpc$ bin of \figref{distance_fraction}, where $\roughly 50 \mhyphen 90 \%$ of satellites are predicted to have undetected tidal tails. Crater II, a classical satellite just below the mass range considered here ($M_* = 10^{5.55} \Msun$) at a distance of 117.5 \kpc \citep{Ji:2021}, also has strong evidence of tidal disruption, including a low velocity dispersion and low surface brightness \citep{Sanders:2018, Fu:2019, Ji:2021, Borukhovetskaya:2022}. \citet{Pace:2022} highlight five additional Milky Way dwarf spheroidals as possibly tidally disrupting. They compare the half-light radius of each satellite to the Jacobi radius at pericenter and find that Bo\"{o}tes I, Bo\"{o}tes III, Grus II, Segue 2, and Tucana IV have densities and pericenters that make them likely candidates for tidal disruption. However, all of these satellites are below the stellar mass range considered in this work. Observations have also provided evidence of extended stellar halos around classical dwarfs, such as Sculptor \citep{Westfall:2006} and Carina \citep{Munoz:2006}. All of these observations suggest that the Milky Way dwarf galaxies may be more tidally disrupted than was originally known, and motivate further searches for evidence of tidal features around known satellites.
% Sats in this mass/dist range: sculptor, Leo I, Fornax, SMC, LMC, and Sgr, Antlia 2.

\subsection{Predictions for Future Detectability}
\label{sec:preds}

\begin{figure}[tp]
    \centering
    \includegraphics[width=0.99\columnwidth]{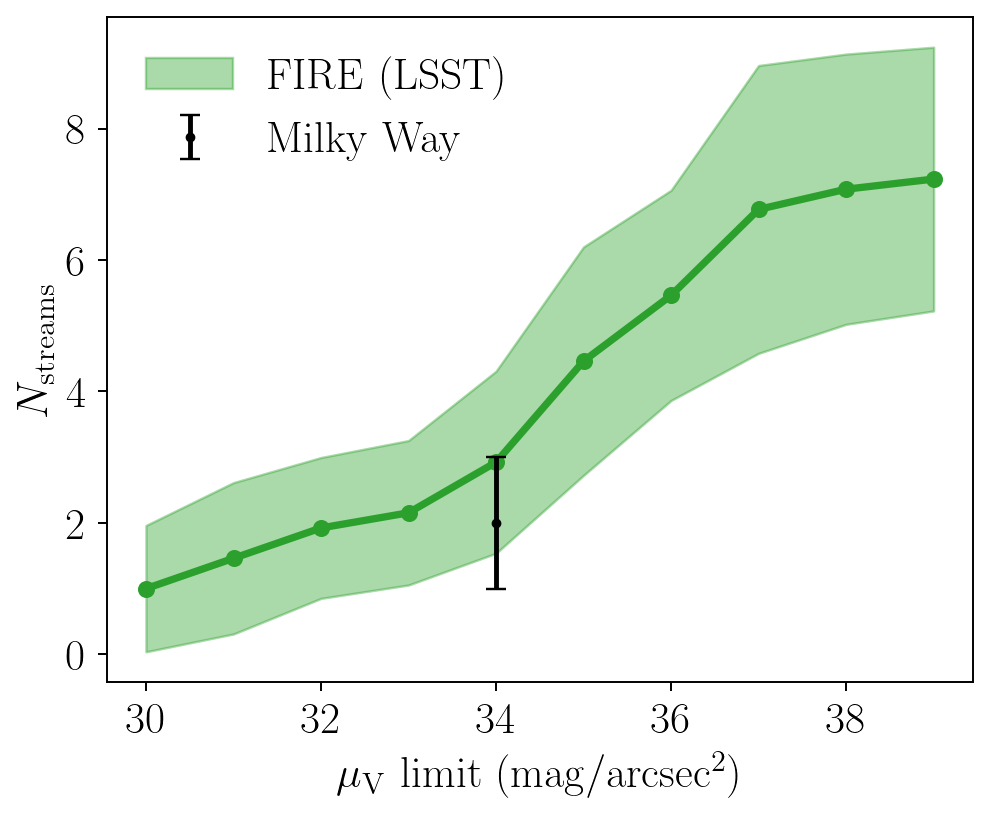}
    \caption{Mean number of detectable high-stellar mass streams in LSST as a function of surface brightness limit. The green curve represents the number of detectable streams with mock LSST observations. The black point is the current number of detectable streams in the Milky Way at the DES surface brightness limit, in this stellar mass range. These results are consistent with previous predictions \citep{Sharma:2012}.}
    \label{fig:rubin_det}
\end{figure}

The Vera C. Rubin Observatory Legacy Survey of Space and Time \citep[LSST;][]{LSST:2009, LSST:2019} will enable the discovery of even more low-surface brightness and distant stellar streams throughout the Milky Way halo. LSST is a photometric survey similar to DES, but covering more of the sky ($18,000 \deg^2$) and extending to fainter magnitudes ($r < 27.5$). In order to estimate the near-future detectability of the FIRE streams, we produce mock LSST observations. We follow the procedure outlined in \secref{mocks}, but sample magnitudes in the LSST $ugrizY$ bands, within the range $17 < r < 27.5$ and using photometric errors estimated from the LSST DESC DC2 simulated sky survey \citep{DC2:2021}.

Due to the difficulty in estimating the stream surface brightness limit in LSST before the availability of real survey data, we estimate the detectable number as a function of surface brightness limit. In \figref{rubin_det}, we show the predicted detectable number of streams as a function of surface brightness limit with LSST photometry. The green curve represents the mean number of detectable streams per halo in the LSST mock observations, and the black point represents the currently detectable number of streams in this stellar mass range ($M_* \gtrsim 5 \times 10^5 \Msun$) in the Milky Way, given the limit of DES. The surface brightness limit for stream detectability with LSST remains uncertain, but will be fainter than that of DES due to the increased survey depth and improved photometric precision. As stream searches are conducted in real LSST data, we will be able to determine detectability limits and thereby the number of streams that should be detected in order to remain consistent with the FIRE simulations. %This prediction is consistent with previous estimates of the detectability of accreted structures with LSST \citep{Sharma:2012}.

If these streams are not detected as we approach fainter surface brightness limits for stream detection, that will indicate an intriguing conflict between the population of tidal remnants in FIRE and in the Milky Way.

\subsection{Inconsistencies Between FIRE and the Milky Way}
\label{sec:inconsistencies}

Despite some similarities, important discrepancies remain between the populations of stellar streams in FIRE and the Milky Way. The pericenters and apocenters of the detectable FIRE streams are larger than those measured for Milky Way streams, on average. Furthermore, the consistency of the number and stellar mass distributions rely on the near-future discovery of a population of high-stellar mass, low-surface brightness streams in upcoming surveys.

\begin{figure}[tp]
    \includegraphics[width=0.99\columnwidth]{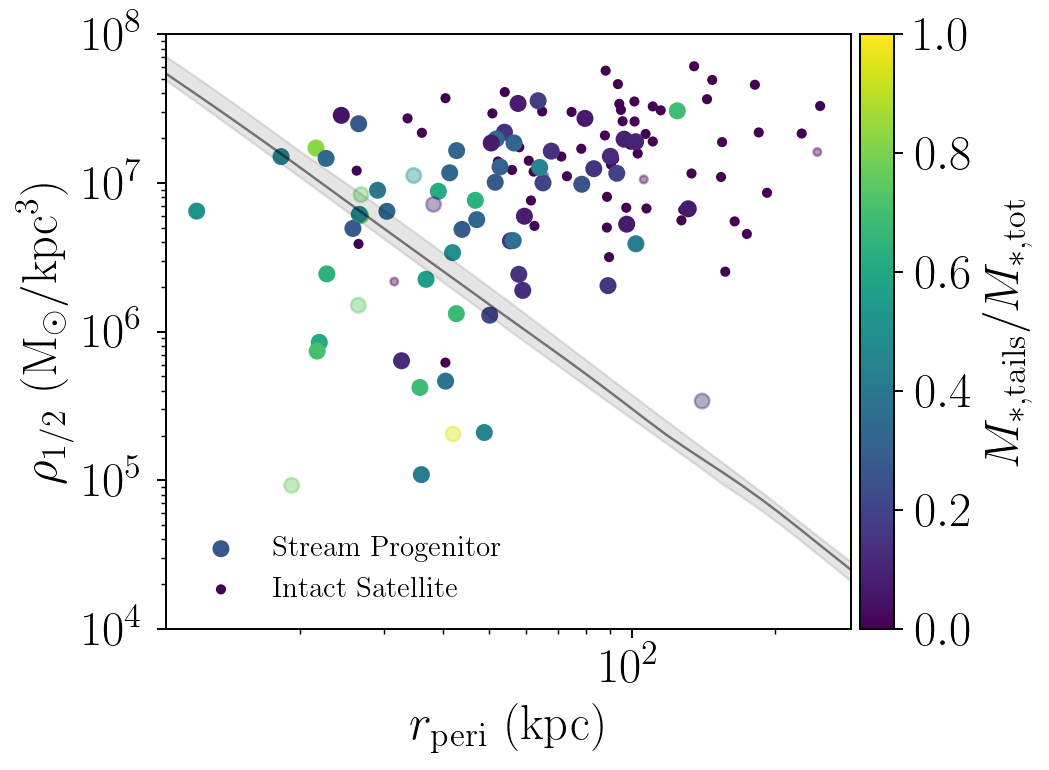}
    \caption{Comparison between the half-light radii and tidal radii of FIRE intact satellites and stream progenitors. The gray line represents twice the average enclosed density at pericenter radius (the shaded band represents the scatter between hosts). Systems with densities below the gray line have tidal radii (Jacobi radii) smaller than their half-light radii and are expected to disrupt, and systems above the gray line have tidal radii larger than their half-light radii and are not expected to disrupt. The points are colored by fraction of stellar mass in their tidal tails. Systems with lower densities have a larger fraction of disrupted material, however many systems are disrupting with large relative densities at pericenter.}
    \label{fig:disrupt}
\end{figure}

The formation of stellar streams at large pericenters, and the phase-mixing of systems on orbits similar to the Milky Way stellar streams (\figref{peri_apo_hists}) may suggest that the FIRE simulations are over-disrupting satellites. More precisely, FIRE may be disrupting and phase-mixing systems too quickly and/or on the wrong orbits. This would have implications for other comparisons between satellite populations in FIRE and the Milky Way. In order to examine why systems are disrupting at large distances, we compare the Jacobi radii and half-light radii of these systems at pericenter. \figref{disrupt} shows the density within the half-light radius of each satellite (intact satellites and stream progenitors) relative to two times the average enclosed density of the host galaxy (gray line). This boundary separates satellites with a Jacobi radius larger (above the gray line) or smaller (below) than their half-light radius. We find that many FIRE satellites are disrupting with relatively high densities. However, while these satellites have relatively small half-light radii, they also have an extended stellar halo and therefore a large fraction of their stellar mass (in some cases up to $40\%$) lies outside their Jacobi radius, even at pericenters of $100 \kpc$. These are the stars that make up their tidal tails at $z=0$.

The sizes of low-mass galaxies are sensitive to numerical effects in simulations. \citet{Binney:2002} found that simulations with two species of particles with a large relative mass ratio may be affected by two-body relaxation. This process leads energy to transfer from the more massive particles (in this case, dark matter) to the less massive particles (stars). This would lead the dark matter to sink to the bottom of the potential well, while the distribution of stars is artificially extended. The effect of this process on galaxy sizes is studied in greater detail in \citet{Ludlow:2019}. If the orbital discrepancy were due solely to resolution effects, we may expect to see a difference between the orbits of the streams from the Latte and ELVIS simulations, due to the factor of two difference in resolution. However, we do not find that the ELVIS streams are more consistent with Milky Way observations. In addition, \citet{Samuel:2020} tested the convergence of the radial profiles of satellites in FIRE and found consistent results across resolutions. Nevertheless, future studies of stream populations in higher resolution cosmological simulations will provide valuable insight into the role of numerical resolution in stream formation. For further discussion of numerical disruption in FIRE, see Santistevan et al. (in prep.).

Tidal disruption of satellites in simulations may also be affected by the implementation of baryonic physics models. The FIRE-2 cosmological simulations of isolated galaxies at $M_{*} \roughly 10^4 \mhyphen 10^9 \Msun$ at high baryonic mass resolution ($30 \mhyphen 500 \Msun$) generally agree well with observed galaxy half-stellar-mass radii \citep{Fitts:2017, Wheeler:2019, Richstein:2022, Shen:2022}. That said, at $M_* \roughly 10^7 \mhyphen 10^{10} \Msun$, FIRE-2 galaxies generally experience bursty star formation, and the stellar feedback during these bursts drives out significant gas mass, which over time puffs out the galaxy sizes \citep{El-Badry:2016, Chan:2018}. While this leads to good agreement with the significant population observed to have large stellar radii (such as ultra-diffuse galaxies), it means that FIRE-2 simulations do not reproduce the most compact (densest) galaxies observed at these masses \citep{Chan:2018, Garrison-Kimmel:2019, Rohr:2022}. In other words, at $M_* \roughly 10^7 \mhyphen 10^{10} \Msun$, the FIRE-2 simulations produce too many galaxies with large stellar sizes, regardless of resolution. At lower masses, ($M_* \roughly 10^4 \mhyphen 10^7 \Msun$), and sufficiently high-resolution ($30 \mhyphen 500 \Msun$), the FIRE-2 simulations form galaxy sizes that agree with observations, but as \citet{Shen:2022} noted, at the necessarily lower resolution of the FIRE Milky Way-mass simulations ($3500 \mhyphen 7100 \Msun$), satellite galaxies at $M_* \lesssim 10^7 \Msun$ are systematically larger than observed. Indeed, a recent re-simulation of one of the Milky Way-mass galaxies at 8 times higher mass resolution ($880 \Msun$) confirms that, in the current simulations at $7100 \Msun$ resolution, satellite galaxy sizes are well-converged at $M_* \gtrsim 10^7 \Msun$. However, below $10^7 \Msun$, the satellite sizes are too large because of limited resolution (Wetzel et al. in prep.). In summary, this combination of physical and numerical effects likely causes the satellite galaxies in the Milky Way-mass simulations we examine here to be larger (on average) than observed and thus may contribute to them disrupting into streams at too-large of Galactocentric distances and/or to phase-mix too quickly once disrupted \citep{Errani:2019}. In addition, the Milky Way analogs simulated in FIRE are known to have later disk-settling times than the Milky Way \citep{Gurvich:2022}, perhaps leading to more chaotic inner galaxies than that of the Milky Way. This in turn may lead satellites with smaller orbital radii to disrupt and ultimately phase-mix more quickly than in our own Galaxy.

Future studies of stream populations in other simulations will help to further uncover the effects of numerical resolution and feedback prescriptions on stream formation, including analysis of streams in the higher-resolution version of m12i (Triple Latte; Wetzel et al. in prep.).
%The Triple Latte simulations are currently being run at 8 times higher resolution than the Latte simulations used in this work (Wetzel et al. in prep.). Comparisons of stream populations in Latte and Triple Latte will enable tests of the resolution effects on satellite disruption.
In addition, other suites of cosmological simulations, such as the Auriga simulations \citep{Auriga:2017}, with independent implementations of feedback physics, will provide important comparisons to the results presented here. \citet{Grand:2021} conducted a comparison of the surviving satellite population in the highest resolution Auriga simulation ($m_{\rm b} = 880 \Msun$, comparable to Triple Latte) to their standard resolution simulations ($m_{\rm b} = 10^4 \mhyphen 10^5 \Msun$) and found that $\roughly 1/6$ of the satellites present at high-resolution are artificially disrupted in the lower resolution simulations. Similar studies of stream populations will provide further insight into the questions raised in this work, and enable stronger tests of satellite disruption in our Galaxy and cosmological simulations.

If the simulations considered in this paper are in fact over-disrupting satellite galaxies, that will have implications for the comparisons of the number and stellar mass distributions of streams as well. However, it is unclear whether correcting for this over-disruption will lead to an increased discrepancy between the numbers of simulated and observed stellar streams. If the source of over-disruption is resolved, some systems that we have counted as streams may be converted to intact satellites and, at the same time, systems that we have classified as phase-mixed may instead form coherent tidal streams. It is therefore necessary to conduct similar studies of stream populations across simulations to study the full extent of the effect of numerical resolution and feedback physics on stellar stream populations.

Another possible explanation is that the Milky Way dwarf galaxy streams considered in this work are in fact the highest-density, most coherent components of largely phase-mixed systems. We find that the systems classified as phase-mixed in FIRE are on orbits more consistent with those measured for the Milky Way stellar streams. Many of these phase-mixed systems do have complex morphologies and include components that, when detected in the Milky Way, could be classified as stellar streams. However, even these high-density regions have relatively low surface brightnesses. In addition, this explanation would not resolve the fact that the FIRE simulations are forming streams at larger pericenters and apocenters than has been observed in the Milky Way. Future observations of low-surface brightness features in the Milky Way's stellar halo will reveal whether the Milky Way dwarf galaxy streams are in fact part of larger extended and diffuse structures that are more consistent with the phase-mixed systems in FIRE.

\section{Conclusion}
\label{sec:conc}

We present the first comparison of detected dwarf galaxy streams around the Milky Way to detectable streams in cosmological simulations. We consider the full population of known Milky Way dwarf galaxy streams and compare them to an analogous population identified around Milky Way-like galaxies in the FIRE simulations.

We produce mock DES observations of the FIRE streams and estimate the detectability of their tidal tails and their progenitors. We find that, when taking into account detectability, the number and stellar mass distribution of these streams is consistent between observations and simulations, resolving the ``too-big-to-fail'' problem in stellar streams proposed by \citet{Li:2022}.

However, the orbital distributions of stellar streams differ between observations and simulations. The Milky Way streams have small apocenters and pericenters relative to those in the FIRE simulations. This discrepancy could be due to properties of the Milky Way, the implementation of feedback physics in FIRE, the resolution of the Latte and ELVIS simulations, or by other discrepancies between the baryonic and dark matter physics that dictates tidal disruption in our Universe and their implementation in the FIRE simulations.

Studies of additional cosmological simulations, with higher resolutions and alternative implementations of baryonic or dark matter physics, will allow us to disentangle simulation effects from new physics. In addition, semi-analytic modeling of stream populations around Milky Way-like hosts may be used to constrain the dependence of the stream population on simulation physics, as well as host halo mass, disk properties, accretion history, and environment. Furthermore, with high-resolution N-body simulations of individual streams, we can test the detailed effects of changes to underlying physics models and host properties on stream disruption.

Future observations of Milky Way stellar streams with surveys such as LSST, as well as the Nancy Grace Roman Space Telescope \citep{Spergel:2013} and Euclid \citep{Laureijs:2011}, will enable further tests of the consistency of the populations of tidal remnants in the Milky Way and the FIRE simulations. In particular, the FIRE simulations predict a population of undetected high-stellar mass, low-surface brightness stellar streams, as well as yet-undetected tidal tails around several known satellite galaxies. Many of these systems are predicted to be detectable in LSST. In addition, these imaging surveys will uncover populations of stellar streams around external galaxies \citep[e.g.][]{Carlin:2016, Pearson:2019} and enable similar tests with a larger sample of galaxies.

This work is the first detailed comparison of stream populations in the Milky Way and cosmological simulations. Additional studies, involving alternative cosmological simulations, semi-analytic modeling, high-resolution N-body simulations, and additional observations will further disentangle the effects of numerics, baryonic physics, and dark matter on the population-level properties of stellar streams.

\section{Acknowledgements}

We thank Sidney Mau and Alex Drlica-Wagner for help providing photometric error estimates for DC2.

This work used the Extreme Science and Engineering Discovery Environment (XSEDE), which is supported by National Science Foundation grant number ACI-1548562. This work used the Stampede-2 allocation number PHY210118 to analyze the simulations.
We generated the FIRE simulations using: XSEDE, supported by NSF grant ACI-1548562; Blue Waters, supported by the NSF; Frontera allocations AST21010 and AST20016, supported by the NSF and TACC; Pleiades, via the NASA HEC program through the NAS Division at Ames Research Center.

NP and RES gratefully acknowledge support from NASA grant 19-ATP19-0068. RES additionally acknowledges support from the Research Corporation through the Scialog Fellows program on Time Domain Astronomy, from NSF grant AST-2007232, and from HST-AR-15809 from the Space Telescope Science Institute (STScI), which is operated by AURA, Inc., under NASA contract NAS5-26555. IBS received support from NASA, through FINESST grant 80NSSC21K1845.
AW received support from: NSF via CAREER award AST-2045928 and grant AST-2107772; NASA ATP grant 80NSSC20K0513; HST grants AR-15809, GO-15902, GO-16273 from STScI.
TSL acknowledges financial support from Natural Sciences and Engineering Research Council
of Canada (NSERC) through grant
RGPIN-2022-04794.

For the purpose of open access, the author has applied a Creative Commons Attribution (CC BY) licence to any Author Accepted Manuscript version arising from this submission. 
%UK open access thing (by Sergey)

\bibliographystyle{aasjournal}
\bibliography{main}

\onecolumngrid
\appendix
\numberwithin{figure}{section}
\numberwithin{table}{section}

\onecolumngrid
\section{Properties of Streams and Satellites in FIRE}
\label{app:sys_props}

Here we report properties of the simulated streams and satellites considered in this work. \tabref{streams} lists the host halo name, the stellar mass, pericenter and apocenter, and minimum surface brightness in DES and LSST of each stellar stream. The stream population and the derivation of these properties is discussed in greater detail in \secref{sims} and \secref{stream_det}. 
\tabref{sats} includes properties of satellites, including the host halo name, structural parameters (half-light radius, ellipticity, position angle), distance, surface brightness, luminosity, and density. Orbital properties of the FIRE satellites will be published in Santistevan et al. (in prep.). In addition, we list whether each satellite is a stream progenitor (i.e. has tidal tails) and whether it is estimated to be detectable in DES, as discussed in \secref{sat_det}. For each table, we display the first ten rows here. The full tables are included as supplemental files.

\newcommand{\streamscaption}{FIRE stream properties.}
\begin{deluxetable*}{l cccccc}[htp]
\tablecolumns{13}
\tablewidth{0pt}
\tabletypesize{\scriptsize}
\tablecaption{ \streamscaption }
\label{tab:streams}
\tablehead{ \colhead{Host} & \colhead{$\log M_{*}$} & \colhead{$r_{\rm peri}$} & \colhead{$r_{\rm apo}$} &\colhead{$\mu_{\rm V, min}$ (DES)} & \colhead{$\mu_{\rm V, min}$ (LSST)}
\\[-0.5em]
 & (\Msun) & (\kpc) & (\kpc) & $(\mathrm{mag} / \mathrm{arcsec}^2)$ & $(\mathrm{mag} / \mathrm{arcsec}^2)$
 }
\startdata
    Romeo &              6.0 &   26.45 & 158.96 &         38 &          37 \\
    Romeo &              8.6 &   41.93 &  65.82 &         30 &          30 \\
    Romeo &              6.6 &   65.00 & 219.77 &         38 &          37 \\
    Romeo &              6.6 &   63.36 & 252.93 &         $>39$ &          37 \\
    Romeo &              5.8 &   17.68 & 157.31 &         39 &          38 \\
    Romeo &              6.3 &   19.43 & 187.39 &         36 &          35 \\
    Romeo &              6.6 &   41.44 & 122.89 &         36 &          36 \\
    Romeo &              6.2 &   21.25 &  68.54 &         36 &          36 \\
    Romeo &              7.4 &   26.90 & 150.56 &         34 &          34 \\
   Juliet &              8.4 &   96.27 & 111.03 &         31 &          31 \\
\enddata
\end{deluxetable*}

\newcommand{\satscaption}{FIRE satellite properties.}
\begin{deluxetable*}{l cccccccccc}
\tablecolumns{13}
\tablewidth{0pt}
\tabletypesize{\scriptsize}
\tablecaption{ \satscaption }
\label{tab:sats}
\tablehead{
\colhead{Host} & \colhead{
RA, Dec Centroid} & \colhead{$r_{1/2}$} & \colhead{$\epsilon$} & \colhead{$\theta$} & \colhead{$d$} & \colhead{$\mu_{\rm V}$} & \colhead{$M_{\rm V}$} & \colhead{$\rho_{\rm 1/2}$} & \colhead{Stream Progenitor} & \colhead{Detectable}
\\[-0.5em]
 & $(\deg)$ & $(\deg)$ &  & $(\deg)$ & $(\kpc)$ & $(\mathrm{mag/arcsec^2})$ & $(\mathrm{mag})$ &  $(\Msun/\kpc^3)$ &
}
\startdata
Romeo &  (61.97, -13.61) &  0.28 & 0.01 &  82.54 & 136.55 & 30.41 &  -8.59 & $2.50 \times 10^7$ &       True &               True \\
Romeo & (103.54, -32.71) &  1.52 & 0.15 & 118.07 &  61.65 & 25.21 & -15.92 & $3.70 \times 10^6$ &       True &               True \\
Romeo &  (96.97, -66.55) &  0.35 & 0.13 & 138.08 & 180.30 & 29.97 & -10.29 & $9.10 \times 10^6$ &       True &               True \\
Romeo &  (109.70, 41.32) &  0.24 & 0.26 &  18.21 & 208.13 & 29.71 & -10.07 & $1.50 \times 10^7$ &       True &               True \\
Romeo &   (57.63, 51.21) &  0.51 & 0.10 & 110.09 &  95.88 & 31.72 &  -7.66 & $1.56 \times 10^7$ &       True &               True \\
Romeo &  (314.97, 42.80) &  0.90 & 0.11 &  75.28 &  60.64 & 29.80 &  -9.71 & $1.26 \times 10^7$ &       True &               True \\
Romeo &  (266.79, 47.80) &  0.64 & 0.17 &  86.22 &  62.64 & 31.29 &  -7.55 & $2.24 \times 10^7$ &       True &               True \\
Romeo &  (247.78, 24.90) &  0.54 & 0.01 & 100.81 & 114.27 & 28.90 & -11.06 & $9.78 \times 10^6$ &       True &               True \\
Romeo &  (79.10, -39.73) &  0.15 & 0.37 &   9.16 & 254.74 & 30.59 &  -8.58 & $2.47 \times 10^7$ &      False &               True \\
Romeo &   (21.18, 48.15) &  0.87 & 0.33 & 167.98 &  93.02 & 25.04 & -15.45 & $5.43 \times 10^6$ &      False &               True \\
\enddata
\end{deluxetable*}

\onecolumngrid
\section{Stream and Satellite Classification}
\label{app:class}

In order to clarify the terms used in this paper, \tabref{class} lists the resulting classification of a stream or satellite system depending on the detectability of its progenitor and tidal tails. For example, a stream with a surviving progenitor would be classified as an intact satellite (with no tidal tails) if only its progenitor were detectable. As discussed in \secref{sat_det}, a ``surviving progenitor'' is a progenitor that is recovered by the halo finder at $z=0$.

\newcommand{\classcaption}{Classification of Streams and Satellites Depending on Detectability}
\begin{deluxetable*}{l ccc}[htp]
\tablecolumns{13}
\tablewidth{0pt}
\tabletypesize{\scriptsize}
\tablecaption{ \classcaption }
\label{tab:class}
\tablehead{ \colhead{All Detectable} & \colhead{Detectable Progenitor Only} & \colhead{Detectable Tidal Tails Only}
\\[-0.5em]
}
\startdata
Stream with surviving progenitor & Intact satellite & Stream without surviving progenitor \\
Stream without surviving progenitor & N/A & Stream without surviving progenitor \\
Intact satellite & Intact satellite & N/A
\enddata
\end{deluxetable*}

\onecolumngrid
\section{Comparison of Orbit Integration Methods}
\label{app:orbit_comp}

\begin{figure*}[tp]
    \centering
    \includegraphics[width=0.52\textwidth]{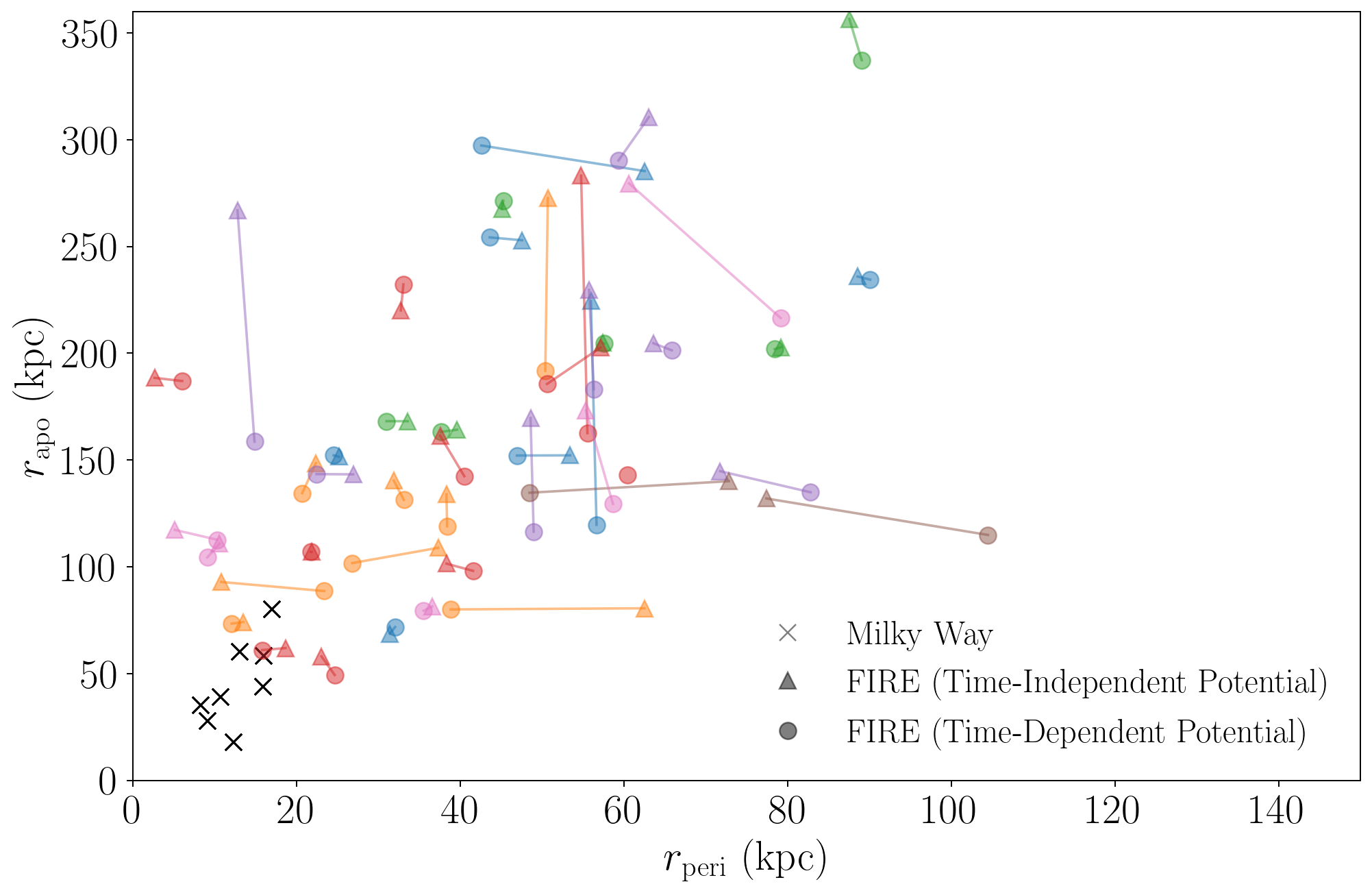}
    \caption{Comparison of pericenters and apocenters of Latte streams computed in time-dependent and time-independent potentials. Each color represents streams belonging to a single host galaxy. The circles represent the values used throughout the paper, which were calculated in a fully time-dependent potential. The triangles represent the pericenters and apocenters of the same streams, calculated by integrating orbits in the time-independent $z=0$ potentials of their host galaxies. This method is analogous to that used to compute the Milky Way stream pericenters and apocenters (black crosses). The majority of streams have a small change in pericenter and apocenter between methods, while a smaller number of streams have more significant changes. These differences are not enough to resolve the discrepancy seen between the orbits of stellar streams in FIRE and the Milky Way.}
    \label{fig:peri_apo_comp}
\end{figure*}

As discussed in \secref{orbits}, the methods used to compute pericenters and apocenters of the FIRE and Milky Way stellar streams are not entirely consistent. Here we discuss the differences between these calculations and the resulting uncertainties. To compute the orbits of the FIRE streams, we trace the positions of the simulated star particles back through the 600 output snapshots, thereby determining the true orbital history in a fully time-dependent potential. However, to compute the pericenters and apocenters of the Milky Way stellar streams, we integrate the orbits of the member stars in a time-independent potential, which is known to introduce biases into derived orbital parameters \citep[][Santisteven et al. in prep.]{D'Souza:2022, Lilleengen:2021}. 
In order to examine the uncertainties on the Milky Way stream orbits due to the assumption of a time-independent potential, we also integrate the orbits of the FIRE streams from the Latte simulations in a time-independent potential. We use the potential models from \citet{Arora:2022}, fit via basis function expansion to the host galaxy potential in each of the Latte simulations at $z=0$. We integrate the orbits using \code{AGAMA} \citep{Vasiliev:2019}. \figref{peri_apo_comp} illustrates the difference in the resulting pericenters and apocenters of streams in the Latte simulations. The circles represent the pericenters and apocenters derived from the fully time-dependent potentials, which are used throughout this paper. The triangles represent the pericenter and apocenter values derived from integrating orbits in the $z=0$ potential. Lines connect the two values for each simulated stream. Each color represents a different Latte host galaxy. The majority of the streams have small changes ($< 10 \%$) in pericenter and apocenter. Some ($\roughly 1$ per host galaxy) have much larger changes. The large differences in either pericenter or apocenter are largely due to artificial fanning of the stream after the first pericenter or apocenter when integrated backwards in the fixed time-independent potential. Regardless of method, we find that the discrepancy remains between the simulated and observed streams. The FIRE streams on average have significantly larger pericenters and apocenters than the Milky Way streams.

\end{document}